\documentclass[pre,aps,amsmath,amssymb,twocolumn,floatfix,preprintnumbers]{revtex4}
\usepackage{graphicx,color}
\usepackage{epsfig}
\usepackage{dcolumn}

\begin{document}

\title{Simulation studies of stratum corneum lipid mixtures}

\author{Chinmay Das}
\affiliation{ School of Physics and Astronomy,
University of Leeds, LS2 9JT, United Kingdom }
\author{Massimo G. Noro}
\affiliation{Unilever R\&D, Port Sunlight, Wirral, CH63 3JW, United Kingdom}
\author{Peter D. Olmsted}
\affiliation{ School of Physics and Astronomy,
University of Leeds, LS2 9JT, United Kingdom }

\begin{abstract}
We present atomistic molecular dynamics results for fully 
hydrated bilayers composed of ceramide NS-24:0, free fatty acid 24:0 
and cholesterol, to address the effect of the different components in 
the stratum corneum (the outermost layer of skin) lipid matrix on its 
structural properties. Bilayers containing ceramide molecules show 
higher in-plane density and hence lower rate of passive transport 
compared to phospholipid bilayers.  At physiological temperatures, 
for all composition ratios explored, the lipids are in a gel phase 
with ordered lipid tails. However, the large asymmetry in the lengths of 
the two tails of the ceramide molecule leads to a fluid like environment 
at the bilayer mid-plane. The lateral pressure profiles show large local 
variations across the bilayer for pure ceramide or any of the two 
component mixtures.  Close to the skin composition ratio, the lateral 
pressure fluctuations are greatly suppressed, the ceramide tails from 
the two leaflets interdigitate significantly, the depression in local 
density at the inter-leaflet region is lowered, and the bilayer have 
lowered elastic moduli. This indicates that the observed composition 
ratio in the stratum corneum lipid layer is responsible for both the 
good barrier properties and the stability of the lipid structure 
against mechanical stresses.  
\end{abstract}


\maketitle
\section{Introduction}
Stratum corneum (SC), the outer layer of the skin
\cite{freinkel.skin.01}, provides the main barrier against water loss
\cite{scheuplein.sc.rev.71} and invasion by foreign pathogens.  During
their life-cycle, cells formed in the basal layer of the epidermis
change their shape and composition. They progressively occupy outer
layers of the epidermis, until they are peeled off from the outer
layer. SC is often viewed as a {\em bricks and mortar} structure
\cite{michaels.sc.brick.75} with corneocytes, the keratin filled
non-viable disc like cells, arranged like {\em bricks} in a lipid
mixture forming the {\em mortar} phase \cite{elias.sc.cryo.77}.  The
three main components of the SC lipid matrix are a family of ceramide
sphingolipids (CER), cholesterol (CHOL) and free fatty acid (FFA)
\cite{norlen.sccomp.99, weerheim.sccomp.01}.  Selective inhibition of
any one of the ceramide \cite{holleran.cer.scbar.91}, cholesterol
\cite{feingold.chol.scbar.90} or FFA \cite{mao-qiang.fa.scbar.93} is
known to compromise the barrier function of the skin. However, how the
three components affect the lipid matrix properties at the molecular
level is not known. To our knowledge, there have been only a few
previous attempts at atomistic modeling of the SC lipid layer.  Most
notably, \citet{holtje.fachol.01} simulated a two component mixture of
fatty acid and cholesterol. \citet{pandit.cer2.06} simulated a bilayer
composed of symmetric CER~NS~16:0 molecules. \citet{notman.dmso.07}
used atomistic simulation to investigate the effect of DMSO molecules
on a hydrated bilayer composed of CER~NS~24:0. But a systematic study
of the effects of the three components is lacking and this study is
specifically concerned with understanding the interplay of the three
components which endows the skin with an almost contradictory
combination of pliability and an extremely high penetration barrier.

There are at least 9 different classes of ceramide found in human SC,
with minor modifications in the head group region and the addition of
an esterified fatty acid in the case of Ceramide 1. All the ceramides are
conspicuous by having a large asymmetry in the length of the two tails
and a large polydispersity in the fatty acid tail lengths
\cite{farwanah.sccomp.cer.05}. Similar polydispersity is found also in
the length of the free fatty acids \cite{norlen.sccomp.ffa.98}.
Realistic representation of such a complex collection of molecules
with atomic details is beyond current computational capabilities.
Instead we choose just one representative ceramide, ceramide NS (also
referred to as ceramide 2), with an asymmetric but monodisperse tail
length.  Ceramide NS is the most abundant species among the ceramide
family. Its fatty acid tail is chosen to be 24:0, guided by the
relative abundance of the different tail lengths in human SC
\cite{farwanah.sccomp.cer.05}; while its sphingosine motif is chosen
to have 18 carbons. Similarly we choose only FFA 24:0 because it is
the most abundant free fatty acid found in SC lipid layer
\cite{norlen.sccomp.ffa.98}. Fig~\ref{fig.scskeletal} shows a skeletal
representation of the molecules.

Between the corneocytes, the lipid matrix shows regular electron
density variations, similar to lipid multilayers. This is not
necessarily the only possible arrangement in the SC lipid matrix. In vitro
experiments show the possibility of asymmetric leaflets
\cite{mcintosh.asym.sc.03} and multiple layer thicknesses, with
indications that ceramide 1 connects different bilayers
\cite{bowstra.sc.cer1.98}. The lipids are predominantly in a gel
phase, possibly a single continuous gel phase \cite{norlen.scgel.01}
or with fluid regions \cite{bouwstra.sc.isgel.02}. In the skin, the
corneocytes are arranged in clusters with the lipid matrix extending
through the full depth of the SC at intervals
\cite{schatzlein.sc.confocal.98}. These regions show a much lower
permeation barrier than the layers between corneocytes
\cite{schatzlein.sc.confocal.98} and it is not known whether or not
the lipids there are arranged in a multilayer structure.  There is a
hydration gradient across the SC, with the average water content
around 30\% by weight \cite{warner.sc.hyd.88, bernstein.sc.hyd.96}. 
How the water molecules
are arranged in the SC is not completely known.

With this uncertainty about the arrangement of the lipids in SC, and
with the computational limitations imposed by detailed atomistic
simulations, we study lipid mixtures arranged in symmetric bilayer
structures in excess water. 
An atomistic potential is chosen in this study to have a
faithful representation of the lipids involved. In the absence of
prior atomistic simulations, or, experimental results on hydrated
SC lipid bilayers, a coarse-grained description is not appropriate
for this system.
The water is required to stabilize the
bilayer structure, and can be viewed as a replacement for the layering
field imposed by the flat corneocytes. The lateral arrangement of the
components and the properties of such a hydrated bilayer will probably 
be representative of the lipid matrix confined between the
corneocytes.

The rest of the paper is organized as follows: In the section
Simulation Details we outline the force field and protocols used for
the simulations, and define the various properties measured. Next we
provide the results for a wide range of compositions, concentrating
mainly at temperature T=340K, which is much higher than the skin
temperature. However, much of the results were found to remain
qualitatively the same even at the skin temperature $\simeq$ 300K.  We
present the results for the higher temperature because we are more
confident of the equilibration at 340K than at skin temperature
T$\simeq$300K. Finally we summarize the main findings and present an
outlook for further studies on SC lipid structures. 

\section{Simulation details}
\subsection{Force field}
The interaction parameters used in the simulation are based on the
united atom OPLS force field \cite{jorgensen.ff.88} with modifications
for the nonpolar CH$_2$/CH$_3$ groups \cite{chiu.ff.95} that
accurately reproduce experimental quantities for lipid molecules
\cite{berger.ff.97}. The polar hydrogens were included explicitly and
the partial charges for the headgroup atoms were selected to conform
with molecules having similar structures and simulated using the same 
force field in
the literature. The dihedral potentials in the hydrocarbon tails were
described by the Ryckaert-Bellemans term \cite{ryckaert.ff.75}.

The force field used for the ceramide molecules has been used
previously \cite{notman.dmso.07}.  The topologies of the polar part of
the fatty acid and the cholesterol are the same as used in
\cite{holtje.fachol.01}.  The water is modeled with the SPC potential
\cite{spcwater}.

\begin{figure}[htbp]
\centerline{\includegraphics[width=3.25in]{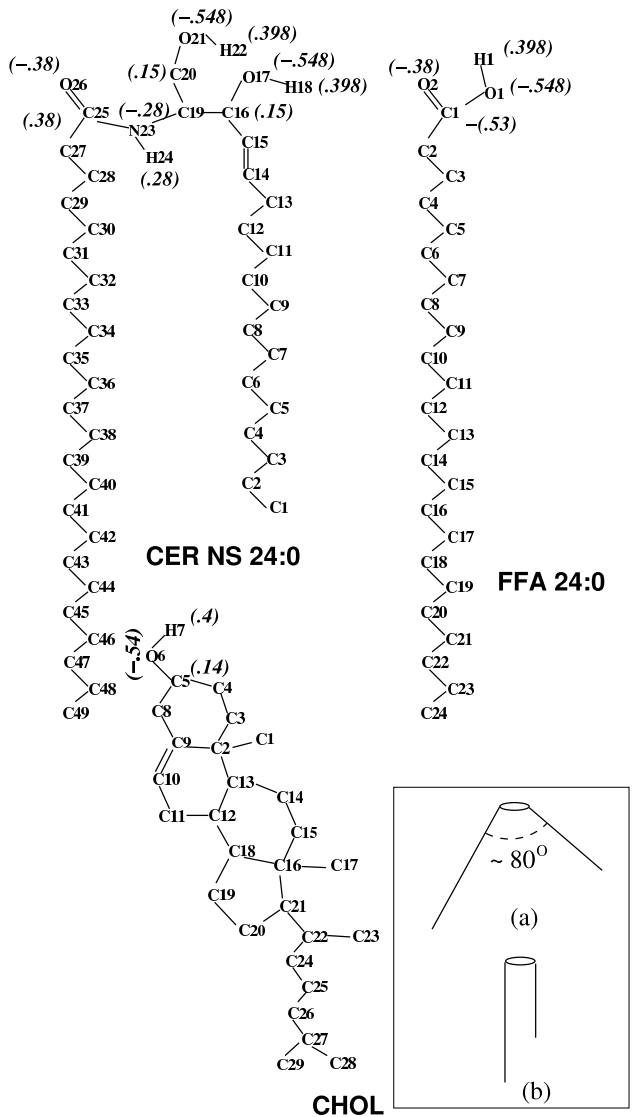}}
\caption{Skeletal representation of the atomic arrangements of the
  ceramide (CER~NS~24:0), the free fatty acid (FFA~24:0) and
  cholesterol (CHOL). Partial charges used in the simulation are
 shown in brackets. Inset: (a) V-shaped and (b) hairpin
  arrangements of the two tails of a CER molecule.
  \cite{dahlen.cer224.xray.79}.}
\label{fig.scskeletal} 
\end{figure}

\subsection{Simulations} 
We used extended ensemble molecular dynamics simulation at constant
temperature and pressure (NPT) ensemble with GROMACS molecular
dynamics package \cite{gromacs95,gromacs05,gromacs.manual}.
Temperature was controlled by Nos\'e-Hoover thermostats
\cite{nose.thermostat.84,hoover.thermostat.85} coupled separately to
the lipid and water molecules with time constant 5ps.  Pressure was
controlled by an anisotropic Parrrinello-Rahman barostat
\cite{parrinello.barostat.81,nose.barostat.83} with time constant 5ps
and compressibility $4.5\times10^{-5}$/bar. The off-diagonal terms of
the compressibility matrix were set to zero to preserve the orthogonal
shape of the simulation box. Standard periodic boundary conditions were
applied in all three directions. A simple group-based cut-off was used for
calculating electrostatic interaction. The cut-off length was chosen
to be $1.2\,\textrm{nm}$ for both the Van~der Waals and electrostatic 
potentials. 
Because the lipid molecules considered in this study do not fully ionize, 
the dominant electrostatic interaction is the dipole-dipole interaction, with
the dipoles made up of the partial changes separated by fixed bond lengths. This
is quite different from the case of phospholipids, where fully ionized phosphate group
introduces a large dipole moment from well separated N$^{-}$ and P$^{+}$ charges.
As a specific example, the C25=O26 atoms on CER molecule
(see fig.~1) form a dipole from equal charges of magnitude $0.28\,\textrm{e}$ separated
by a fixed distance of $0.123\,\textrm{nm}$. This creates a dipole moment of $1.65\,\textrm{D}$,
which is roughly $15$ times smaller than the dipole moment at the head groups in DPPC
bilayer \cite{wohlert.04}. Noting that the dipolar interaction falls of as $1/r^3$ and
that for DPPC bilayer effect of electrostatic interaction with the periodic images become
negligible for system sizes $\sim 1024$ lipid molecules \cite{wohlert.04}, we estimate that for 
the dipole moments involved in our system, the effect of electrostatic interaction with the 
periodic images
become neglibile for systems containing $\sim 168$ lipids. We have carried out 
simulations with pure CER molecules and 2:2:1 composition system both with group-based
cut-off and particle mesh Ewald summation (PME) for several system sizes. The results
are presented in the appendix and validates the
independence of the results presented on long range electrostatic interaction and
system size. Because the use of PME requires more than twice the computational 
cost as compared to using cut-off and it has no
effect on the validity of the results presented here, we restrict ourselves to group-based
cut-off for handling electrostatic interaction in the simulations presented in the main paper.
 
All bond lengths were
constrained using the SHAKE algorithm \cite{ryckaert.constraint.77}.
The analytic SETTLE algorithm was used to handle the rigid SPC water
molecules \cite{miyamoto.settle.92}. We used a timestep of $2\,\textrm{fs}$ 
for $\textrm{T} \le 340\,\textrm{K}$ and $1\,\textrm{fs}$ for high temperature simulations.
During the the production run, configurations at intervals of $0.5\,\textrm{ps}$ 
were stored for further analysis.

The SC lipid matrix (in mice) shows a pH of 
$\sim 6$ \cite{hanson.scph.02}. The ionization state of FFA is
known to modify the structure of CHOL-FFA (C16:0) mixtures 
\cite{ouimet.pachol.03}. For ceramide-cholesterol mixtures containing
palmitic acid (C16:0) and oleic acid (C18:1), reported values of the
pK$_{\textrm{a}}$ lie  in the range of 6.2-7.3 \cite{lieckfeldt.scph.95}.
Typically longer tailed FFA are expected to have higher
pK$_{\textrm{a}}$; \textit{e.g.},
for  hexacosanoic acid (C26:0) in an egg-phosphatidylcholine bilayer
pK$_{\textrm{a}}\sim 7.4$. This suggests that a fraction of the FFA
should ionize at  skin pH. 
However, the experimental
evidence for the effect of pH on SC lipid mixture is not clear. 
Surface force  and AFM measurements on SC lipid mixtures 
containing free fatty acid seem to show no detectable effect in a wide 
range of pH (between 3.0 and 7.0) \cite{norlen.sc.afm.07}. The apparent
insensitivity to pH for SC lipid mixture may be because the bilayers
containing long tail ceramides remain in a dense gel phase and do not
allow much structural freedom to the FFA molecules. With this background,
we chose to simulate only {\em un-ionized} FFA - not having free charges
allow fast calculation of the forces with cut-off as alluded before. 

\subsection{Initial structures}
In all but one of the six different crystal structures formed under
varying conditions, the sphingosine and the fatty acid tails of
CER~NS~24:0 arrange themselves with a large opening angle V-shaped
structure \cite{dahlen.cer224.xray.79}.  Only low temperature
crystallization from solution leads to a hairpin arrangement of the
tails (inset of Fig.~\ref{fig.scskeletal}).  To ensure a hairpin arrangement of
the ceramides, we start with a multilayer anhydride system with the
ceramides placed in a slightly expanded hexagonal lattice with a hairpin
structure at 300K, and relax the configuration with a 20\,ns
NPT MD run.  A bilayer from the multilayer was then placed in a larger
box and the box was filled with water. We ran the system for 5\,ns with
the ceramide molecules fixed (to hydrate the bilayer properly), and
then for 2\,ns with only the terminal methyl group on all the lipid
tails frozen, so that the rest of the molecules reorient themselves to
accommodate the water environment. This configuration was used as the
initial configuration for the pure CER system. A further 10\,ns NPT MD
run was used at each temperature before the production runs.

\begin{table}[htbp]
  \begin{center}
    \begin{tabular}{|c|l|l|l|}\hline
      composition & \multicolumn{3}{|c|}{number of molecules} \\ \hline
      CER:CHOL:FFA & && \\
      (molar ratio) & $n_{\text{CER}}$ & $ n_{\text{CHOL}}$ & $ n_{\text{FFA}}$ \\ \hline
      1:0:0 & 128 & 0 & 0  \\ 
      0:1:0 & 0 & 128 & 0 \\
      0:0:1 & 0 & 0 & 512 \\ \hline
      7:1:0 & 112 & 16 & 0 \\
      3:1:0 & 96 & 32 & 0 \\
      2:1:0 & 86 & 42 & 0 \\
      1:1:0 & 64 & 64 & 0 \\ \hline 
      7:0:1 & 120 & 0 & 16 \\
      3:0:1 & 110 & 0 & 36 \\
      2:0:1 & 102 & 0 & 52 \\
      1:0:1 & 86 & 0 & 84  \\ \hline 
      1:1:1 & 52 & 50 & 52 \\
      1:2:1 & 32 & 64 & 32 \\
      2:1:1 & 64 & 32 & 32 \\
      5:5:1 & 60 & 60 & 12 \\
      2:2:1 & 56 & 56 & 32 \\ \hline 
    \end{tabular}
    \caption{Compositions (as molar ratios) and number of particles
      simulated in this study.  The pure FFA system had 9020 water
      molecules. For all other compositions, there were 5250 water
      molecules. \label{table.composition}}
  \end{center}
\end{table}

For mixtures containing CER molecules, we transform the required
number of CER molecules, chosen at random but maintaining the
same composition in the two leaflets, to either CHOL or FFA.  To convert to FFA
molecules we simply separate the two tails of CER as two separate
molecules and slowly grow the shorter chain derived from the
sphingosine motif to the required length. FFA molecules are 
significantly more mobile as compared to CER molecules. Thus, although
in the initial configuration the two FFA molecules generated from the
same CER molecule are next to each other, they do not introduce
any artificial correlation after the equilibration step.
To convert to CHOL we map
certain atoms of the CER to CHOL. Repeated short NVT MD runs were
performed while keeping the CHOL molecules frozen, and after each such
short run, the atoms of the CHOL were displaced by a small amount
until they reached the equilibrium positions on the molecule.  The
mixtures were adequately equilibrated, typically by energy
minimization followed by a series of short NVT simulations and finally
with at least 10\,ns of an NPT MD run.  
To find a typical relaxation time, we quenched a CER bilayer
from $360\,\textrm{K}$ to $300\,\textrm{K}$ and found that
the area/lipid and the bilayer thickness re-equilibrate to
$300\,\textrm{K}$ values in $\sim 1 \,\textrm{ns}$ time-scale.
For pure FFA or CHOL bilayers,
we simulated multilayers and placed one bilayer in water in the same
way as in the case of CER.
Table~\ref{table.composition} shows
the number of molecules used, as well as the corresponding molar ratios, 
rounded to closest integer ratios. 
Some of the composition ratios explored in this study may not be
achievable in experiments as hydrated bilayers. However, once prepared,
the bilayers are kinetically stable and allow us to
isolate the effects of the different components by studying the
extreme compositions.
In the rest of the paper the different
compositions are referred to by the corresponding molar ratios.

\section{Measured quantities}
In this section we define the different quantities measured from the
simulations. For all composition ratios and temperatures, we use
averages over 10\,ns runs. We estimate the error bars for the measured
quantities from the variance of intermediate averages over 2\,ns
windows.

We define the bilayer normal direction to be the $z$ direction. The
orthogonal box shape along with the small system size ensures that
this direction remains the same as the $z$ direction of the
simulation box. All $z$-dependent quantities reported are averaged
over the lateral direction ($x-y$ plane) in the entire simulation box.

\subsubsection{Structural properties}

At the molecular length scale, the water-lipid interface has a finite
width.  To assign an unique value to the bilayer thickness $2 d$, we
calculate the density of the water molecules as a function of $z$ and
define the position of the interface between the lipid and water as
the $z$ at which the water density decays to $\frac{1}{e}$ of the bulk
water density. We use this criteria, as opposed to the more usual
criteria of the Gibbs dividing surface, because we consider
multi-component lipids in this study and the different components 
have significantly different densities at the boundary. Concentrating on
the approximately exponentially decaying water density gives a simple
and unique method to assign a value to the bilayer thickness.
The average lipid layer density $\bar{\rho}_L$ is
computed by assuming that all the lipid mass is homogeneously
distributed between the two lipid-water interfaces. 
The local density of the lipid molecules
depends on $z$ with a minimum $\rho_L^{min}$ between the two leaflets.

The asymmetry in the lengths of the two hydrocarbon tails of CER can
lead to significant interdigitation.  We define the following 
dimensionless overlap
parameter as a quantitative measure of interdigitation,
\begin{equation}
\rho_{ov} (z) = 4 \frac{\rho_{t} (z) \times \rho_{b}(z)}{\left[ \rho_{t} (z) + \rho_{b} (z) \right]^2 },
\label{eq.overlap}
\end{equation}
where $\rho_{t} (z)$ and $\rho_{b} (z)$ are the densities at $z$ for
the CER from the top and bottom layers respectively. $\rho_{ov} (z) =
1$ if half of the density at $z$ is from the top layer and the other
half is from the bottom layer CER molecules. If only the top or bottom
layer of CER is at $z$, $\rho_{ov} (z) = 0$.  Integrating over $z$, we
define a single length scale $\lambda_{ov} \equiv \int_0^L
\rho_{ov}(z) dz $ to compare the amount of interdigitation for different
compositions. The integration is carried over the whole box, since if
no CER is present, then $\rho_{ov} (z) = 0$ and there is no
contribution to $\lambda_{ov}$.

The area compressibility $\kappa_A$ of the bilayer is related to the
area fluctuation in the NPT ensemble by \cite{allen-tildesley.87}
\begin{equation}
\kappa_A = k_B T \frac{<A>}{<A^2> - <A>^2},
\end{equation}
where $k_B$ is the Boltzmann's constant and T is the temperature. The
angular brackets refer to averages over time for the area $A$ and
square of the area $A^2$.

The system sizes investigated in this work are too small to calculate
the bending modulus by analyzing the undulation modes \cite{evans.90,
 lindahl.00}.  For fully saturated fluid membranes, the bending
 modulus $\kappa$ is approximately related to the area compressibility
 modulus $\kappa_A$ and the thickness of the hydrocarbon tail region $2
 d_c$ through
 \begin{equation}
 \kappa = \frac{\kappa_A (2 d_c)^2}{c_e},
 \end{equation}
 where the constant $c_e$ is estimated to be $24$ from a theory based
 on polymer brushes \cite{rawicz.polybrush.00}. This simple
 prescription is found to work well for a number of different lipid
 bilayers \cite{rawicz.polybrush.00, bermudez.04}. In the present
 simulation, the head groups of the lipid molecules are much smaller
 than that of phospholipids and we use the bilayer thickness $2 d$ as a
 good estimate of the hydrocarbon thickness $2 d_c$. 

\subsubsection{Tail order parameter}
The orientational (nematic) order of the tails is probed through an
order parameter $P_2$, defined by the largest eigenvalue of the second
rank tensor
\begin{equation}
Q_{\alpha \beta} =  \left< \left(
\frac{3}{2} \hat{u}_{i\alpha} \hat{u}_{i \beta} - \frac{1}{2} \delta_{\alpha \beta} \right) \right>,
\end{equation}
where $\hat{u}_{i \alpha}$ is the cartesian component $\alpha$ of some
specific orientation vector on the lipid molecule $i$, and the $<~>$
denotes average over time and the lipid molecules.  For fluid bilayers,
$P_2$ is related
to the NMR deuterium order parameter $S_{CD}$ through the relation
$P_2 = - 2 S_{CD}$ \cite{seelig.74}.  The eigenvector corresponding to
the largest eigenvalue of $Q$ defines the average orientation.
Different carbon atoms along the tail are expected to have different
amount of ordering. To probe the local ordering, we define a local
orientation $P_2(n)$ at carbon atom $n$ on the lipid molecule $i$ 
by the vector direction
$\hat{u}_{i \alpha} (n)$ between carbon atoms $n-1$ and $n+1$.

\subsubsection{Local pressure}
The anisotropic arrangement of lipid molecules in a bilayer leads to
an anisotropic local pressure profile, and hence local stress
variations.  While macroscopic pressure is a well defined quantity in
molecular simulations, its microscopic description is not unique
\cite{schofield.localp.82}. We use the formalism of
\cite{lindahl.localp.00} to define the local pressure tensor at a
given height $z$ as
\begin{equation}
{\mathbf P} (z) = \frac{1}{V_{slice}} \left< \sum_{i \in slice} m_i {\mathbf v}_i
\otimes {\mathbf v}_i - \sum_{i<j}
{\mathbf F}_{ij} \otimes {\mathbf r}_{ij} f(z,z_i,z_j) \right>.
\label{eq.localp}
\end{equation}
Here, ${\mathbf v}_i$ is the velocity of particle $i$, ${\mathbf
  F}_{ij}$ is the force on particle $i$ due to particle $j$ and
${\mathbf r}_{ij}$ is the relative position vector of particle $i$
from particle $j$.  $V_{slice}$ is the volume of a thin slice with
thickness $\Delta z$ and $<~>$ denotes an average over time.  The
first sum runs over the particles in the slice centered at $z$. The
function $f$ determines the contribution from the virial (the second
sum) to the current slice.  $f$ is unity when both particles are in
the current slice. If one or both of the particles are outside the
slice (but the shortest distance between the two particles goes
through the slice); $f$ is respectively chosen to be $\Delta z/|z_i -
z_j|$ and $\Delta z/\left(2 |z_i - z_j| \right)$.  Because constraints
in the simulation (SHAKE algorithm) transfer some of the kinetic
contributions to the constraint potential, we need to consider both
the sums in Eq.~\ref{eq.localp} explicitly.  We calculate the local
pressure profiles by starting with stored configurations separated by
2\,ns and re-evolving the configurations for 200\,ps.

Of particular interest for lipid bilayers is the difference between
the lateral and the normal pressure $\delta P (z) = P_{\text{LAT}}(z)
- P_{zz} (z)$, where $P_{\text{LAT}} (z) = \frac{1}{2} \left[P_{xx}
  (z) + P_{yy} (z) \right]$. The surface tension $\gamma$ of the bilayer is
related to $\delta P (z)$ through \cite{lindahl.localp.00}
\begin{equation}
\gamma = - \int_{-d}^{d} \delta P (z) dz.
\end{equation}
Anisotropic box fluctuations ensure that the bulk water is isotropic,
i.e.\ the pressure tensor has equal diagonal components and zero
off-diagonal components.  This in turn leads to a zero average surface
tension for the lipid bilayer. But locally $\delta P (z)$ goes through
a number of maxima and minima. We define a microscopic stress by
\begin{equation}
\bar{\epsilon}_P = \frac{1}{2 d} \int_{-d}^{ d}  dz \left[ \left\langle\left(\delta
    P(z) - \langle\delta P(z)\rangle \right)^2 \right\rangle \right]^{\frac{1}{2}}. 
\label{eq.localp.en}
\end{equation}
Here, the integration is over the bilayer thickness and the angular
brackets reflect time averages.

\section{Results and Discussion}

\subsection{Pure Ceramide bilayers}

\begin{figure}[htbp]
\centerline{\includegraphics[width=0.45\linewidth,height=2.5in,clip=]{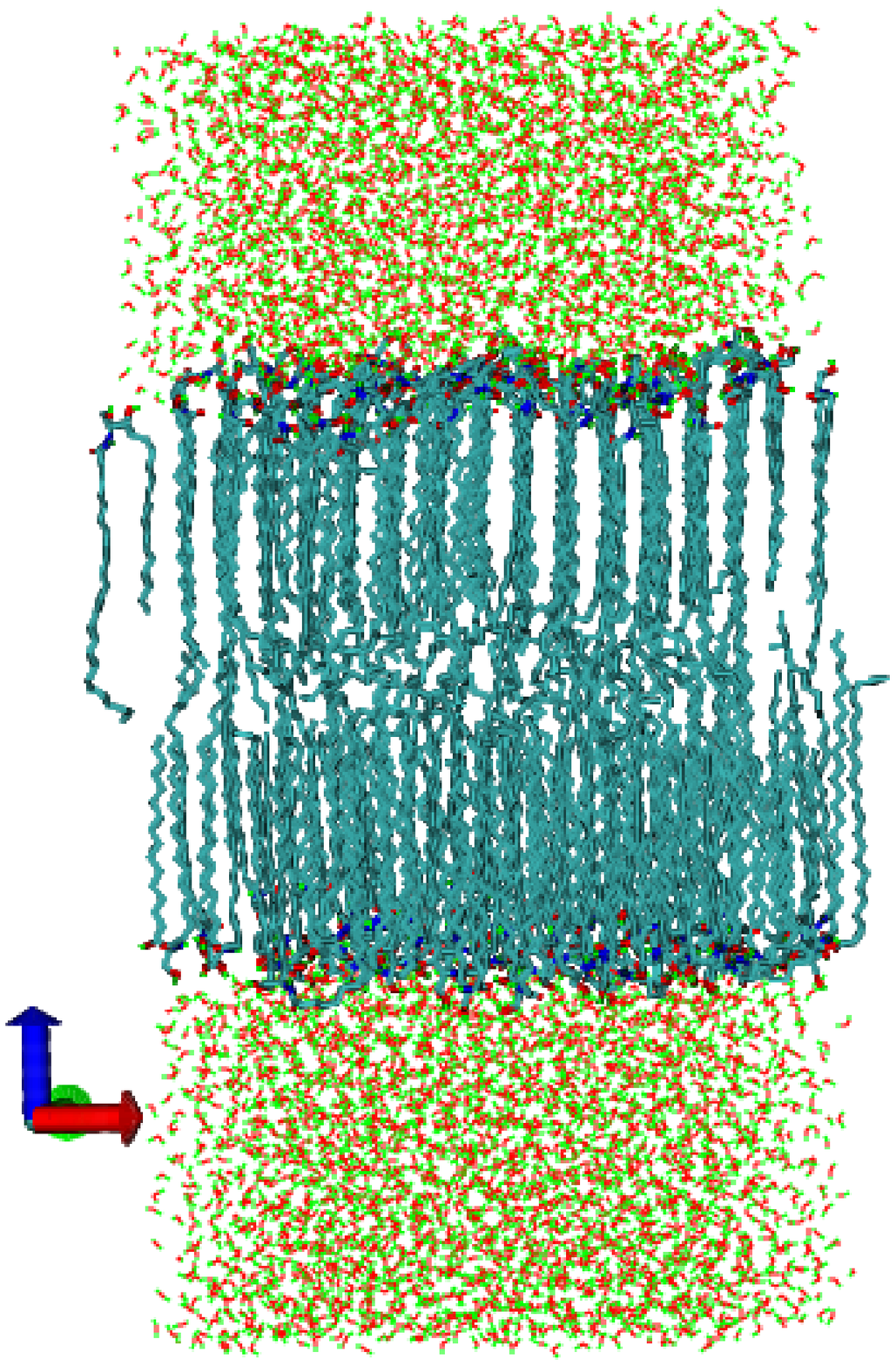} \hfil %
\includegraphics[width=0.45\linewidth,height=2.5in,clip=]{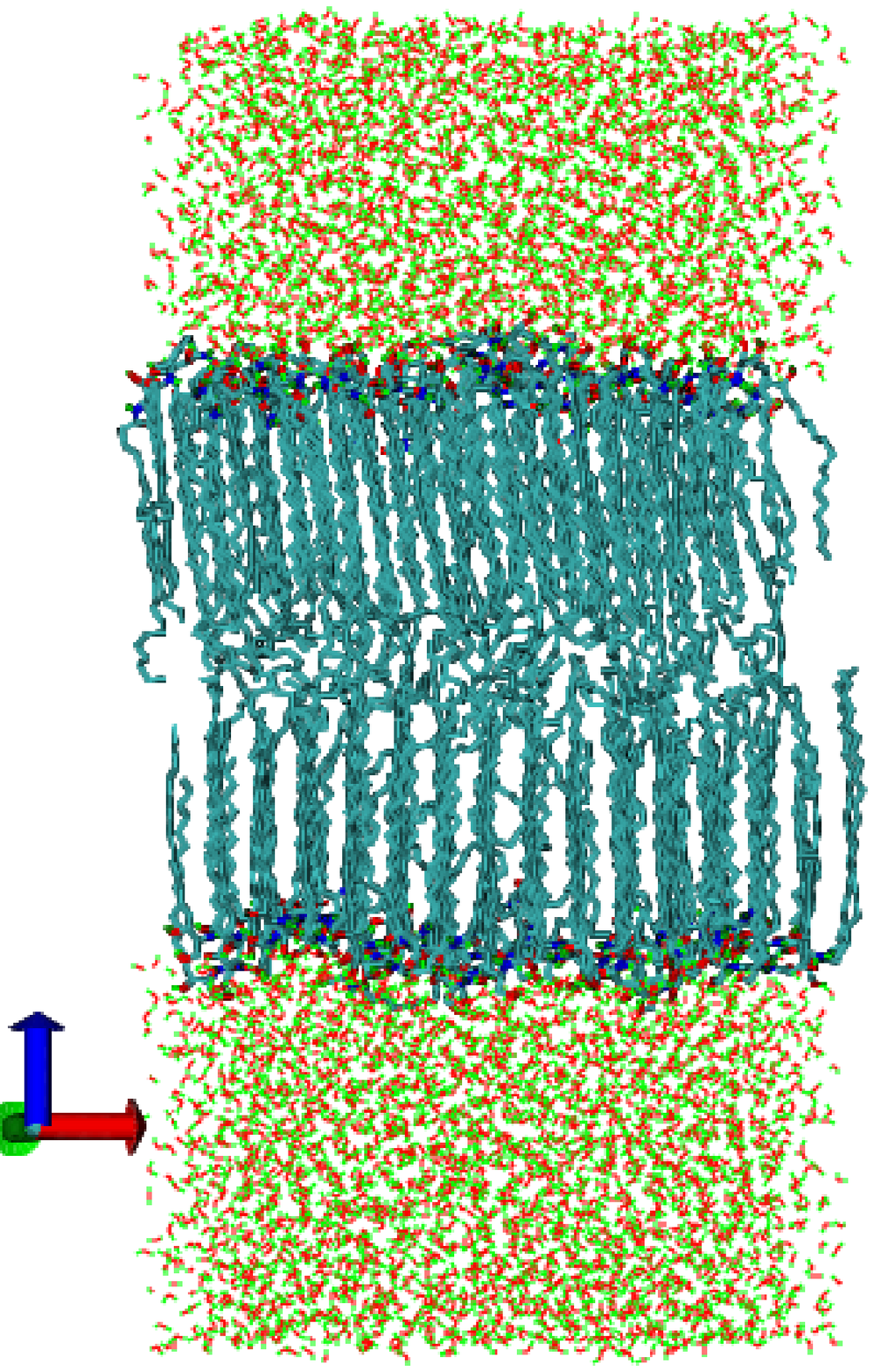} }
\vskip0.2cm
\centerline{(a) \hskip0.6\linewidth (b)}
\vskip0.2cm
\centerline{\includegraphics[width=0.85\linewidth,clip=]{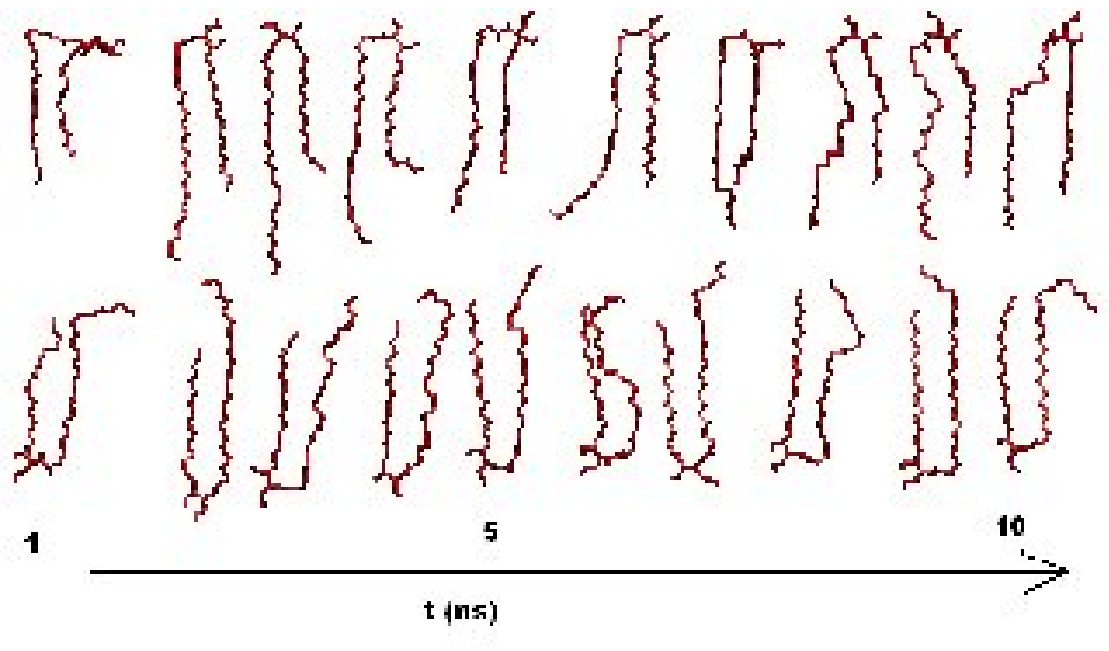}}
\vskip0.2cm
\centerline{(c)} 
\caption{Snapshots of a CER2 bilayer at (a) 300K and (b) 360K. 
While both the leaflets have similar tail ordering, projection in
two dimensions artificially accentuates order in one leaflet at the cost
of showing less order in the other leaflet.
(c) Time trace
of two CER molecules. The perspective for the two molecules are chosen 
independently and each time frame has been shifted arbitrarily.}
\label{fig.snapshot.cer2}
\end{figure}

\begin{figure}[htbp]
\centerline{\includegraphics[width=3.25in,clip=]{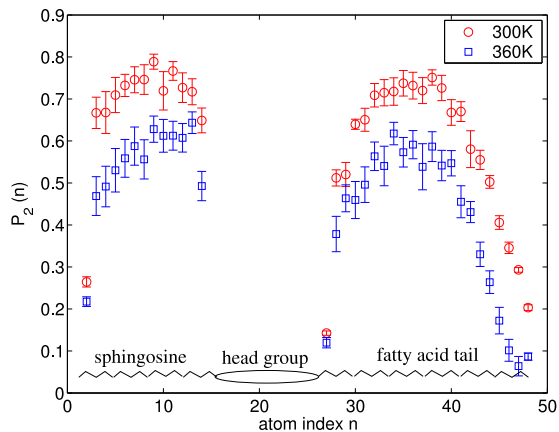} }
\caption{Tail order parameter of CER atoms as a function of atom index
  $n$ (see Fig.~\ref{fig.scskeletal}). 
}
\label{fig.cr2.tailop}
\end{figure}

\begin{figure}[htbp]
\includegraphics[width=3.25in]{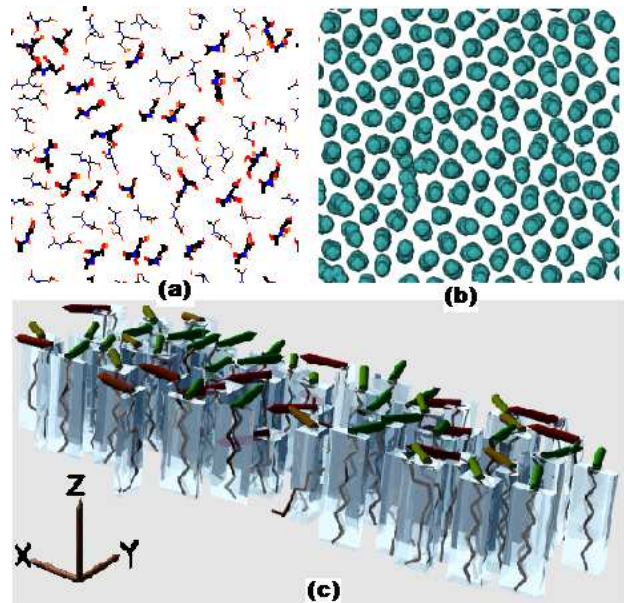} 
\caption{(a) Top view of CER molecule head group arrangement on one
  leaflet at $T=340\,\textrm{K}$. The molecules are drawn in thick or 
  thin lines respectively,
  depending on whether the tangent of the angle between the line
  joining the end-atoms of the head group and the x axis is positive
  or negative. 
 (b) Cross section of CER tail groups a few atoms
  below the head groups, showing local hexatic order. 
 (c) Side view of part of the lipid tails with (orthogonal distance regression) 
 planes containing the tail atoms indicated by transparent boxes. The normals to these
 planes are indicated by solid rods on top of the boxes.
 In a crystalline configuration, the alignment of these planes is regular 
 throughout the sample, while in the rotator phase, there are no correlations 
 between the planes. In the present case, local correlation
 between the hydrocarbon planes persist but allows slow rearrangement 
 because the correlation is not perfect.
  }
\label{fig.cr2head}
\end{figure}

Fig.~\ref{fig.snapshot.cer2} shows snapshots of the CER bilayer at (a)
300K and at (b) 360K.  The tails of the CER align along the $z$
direction with strong nematic ordering. The terminal methyl groups at
the bilayer midplane show a liquid-like disordered arrangement.  With
increasing temperature, the disordered region at the bilayer midplane
gradually thickens. The tail order parameters at 300K and
360K are shown in Fig.~\ref{fig.cr2.tailop}. The terminal groups
(smallest and largest atom indices) have small ordering.  $P_2$ also
decreases close to the head group (atom indices 16 and 25). The C-C
bonds close to the head group orient at an angle to the vertical
direction. Fig.~\ref{fig.cr2head} (a) shows a top view of one leaflet.
The figure shows that the molecules arrange in a zigzag fashion with rows of
molecules having the head group arrangement in alternate 
(orthogonal) directions.
Thus the tail carbon bonds close to the head group point in 
one of these two orthogonal directions, reducing the overall 
average of $P_2$.

In the well-ordered
layer immediately beneath the head groups, the tails have fairly
well-defined hexatic order with very few defects (Fig.~\ref{fig.cr2head}b). 
Moreover, the 
orientations of
the planes containing the lipid tails (Fig.~\ref{fig.cr2head} c )
suggests that
the phase of this ordered layer is intermediate between crystal and
rotator phases (familiar from studies of alkanes of different lengths
\cite{sirota1993rpn}).
While the two dimensional slices show strong sixfold order,
the molecules undergo relatively fast slithering motion of the
tails (Fig.~\ref{fig.snapshot.cer2} c). The coupling between the two leaflets is weak, allowing
the two leaflets to oscillate about each other by more than the inter-lipid
distance in roughly $10\,\textrm{ns}$. Diffusion of the center of masses
of the molecules in the $x-y$ plane is much slower because they can
only diffuse at the defect sites.

\begin{figure}[htbp]
\centerline{\includegraphics[width=3.5in,clip=]{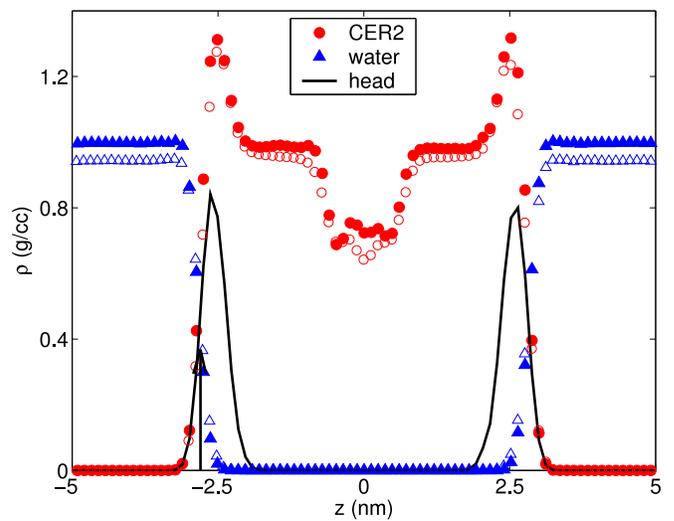}}
\caption{Density of CER atoms (circles), water molecules (triangles)
  for CER bilayer at 300K (filled symbols) and 360K (open symbol). The
  density of head group atoms at 300K is shown with a solid line.}
\label{fig.cr2.lden}
\end{figure}

Fig.~\ref{fig.cr2.lden} shows the density of the CER and water
molecules across the bilayer at 300K (open symbols) and 360K (closed
symbols). The CER density shows a peak at water-lipid interface from
the close packing of head group atoms. There is an almost constant
density shoulder from the dense packing of the hydrocarbon tail atoms,
followed by a dip at the bilayer midplane, which corresponds to the
amorphous inner layer due to the asymmetric ceramide tails. The arrow
indicates where the water density falls below $1/e$ of the bulk water
density at 300K, which is considered to be the lipid-water interface
to calculate the bilayer thickness.

\begin{figure}[htbp]
\centerline{\includegraphics[width=3.5in,clip=]{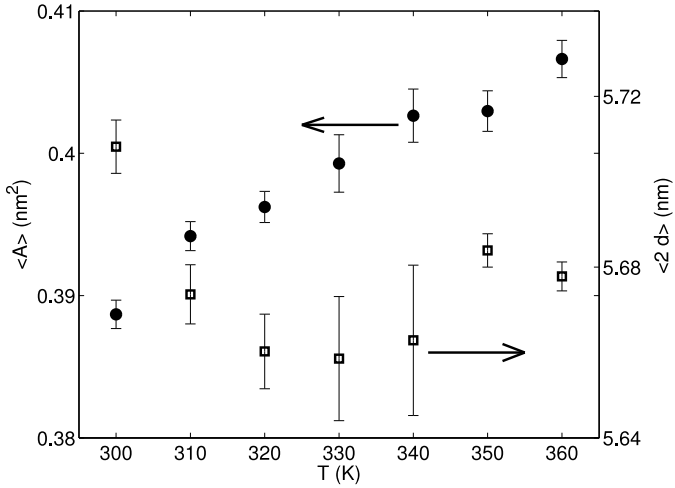} }
\caption{Area/lipid (circles, left y-axis) and bilayer thickness
  (squares, right y-axis) of CER molecules as a function of
  temperature.}
\label{fig.cr2.area}
\end{figure}

\begin{figure}[htbp]
\centerline{\includegraphics[width=3.5in,clip=]{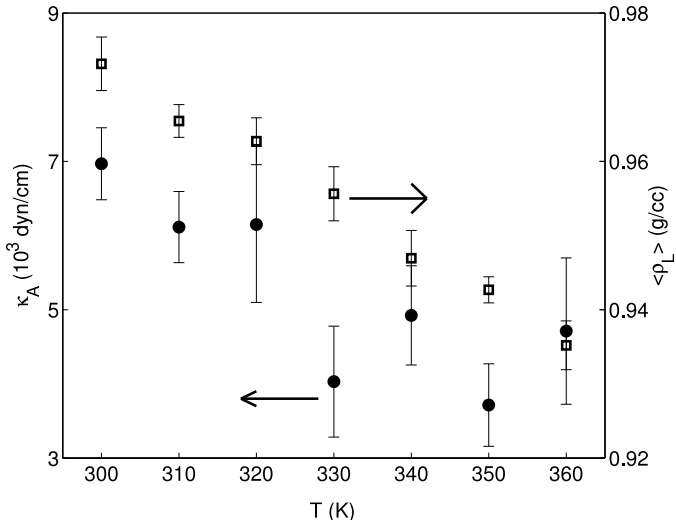} }
\caption{Area compressibility (circles, left y-axis) and average
  density (squares, right y-axis) of CER bilayers as a function of
  temperature.}
\label{fig.cr2.density}
\end{figure}

Fig.~\ref{fig.cr2.area} shows the area per lipid and bilayer thickness
as a function of temperature. The area/lipid at 300K is 0.389~nm$^2$,
which is close to the X-ray result $\sim$~0.4~nm$^2$ for CER bilayer
with molecules in hairpin arrangement \cite{dahlen.cer224.xray.79}.
With increasing temperature, the area/lipid gradually increases, while
the increasingly disordered terminal tail atoms make the thickness
decrease.  The average density and area compressibility
(Fig.~\ref{fig.cr2.density}) both decrease with temperature. Even at
360K, the area compressibility is an order of magnitude larger
($\sim$~4000 dyne/cm) than in most phospholipid fluid bilayers. The
results do not show a sharp transition. Instead the disordered
inter-leaflet region expands smoothly with temperature. The long
chains involved in CER probably will lead to a broad gradual softening
like a waxy material, instead of a sharp transition. Experimentally,
main transition temperatures in ceramide systems have been found, for
multilayer systems, to be of order $394-420\,\textrm{K}$
\cite{dahlen.cer224.xray.79}. One expects a lower temperature for a
single hydrated membrane in solution.

\subsection{Mixed SC lipid bilayers}

\begin{figure}[htbp]
\centerline{\includegraphics[width=3.5in,clip=]{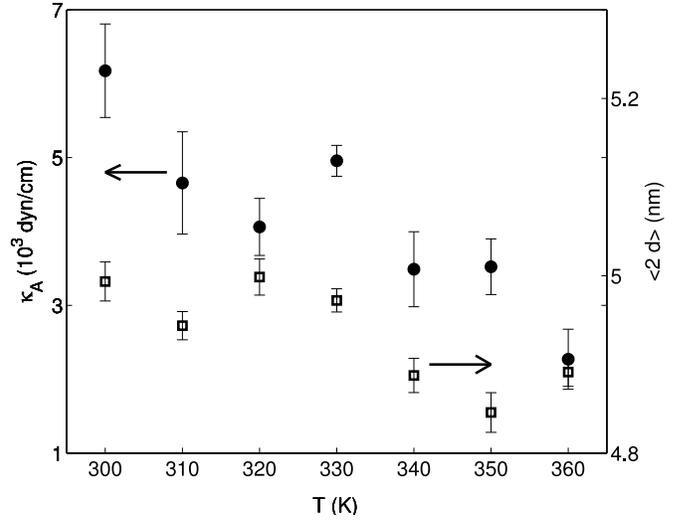} }
\caption{Area compressibility (circles, left y-axis) and average
  density (squares, right y-axis) of a mixed system,
  CER:CHOL:FFA=2:2:1,  as a function of
  temperature.}
\label{fig.cr2.mixed}
\end{figure}

For all the composition ratios investigated 
the temperature behavior in
the simulated range of temperatures is gradual and similar to that of
CER system.
(Fig.~\ref{fig.cr2.mixed} shows results for the composition
 CER:CHOL:FFA=2:2:1). 
For presentation of the results in this section,
we concentrate on 340K.  
A summary of the different measured
quantities is included in Table~\ref{tab.struct}.

\begin{table*}
\begin{center}
    \begin{tabular}{|c|l|l|l|l|l|l|l|}\hline
      composition   & $2 d$ &  $\bar{\rho}_L$ &  $\rho_L^{mid}$ & $\lambda_{ov}$  & $\kappa_A$ ($10^3$) & $\kappa$  ($10^{-11}$)& 
      $\bar{\epsilon}_P$ \\ \hline
      (molar ratio) & (nm)  &   (g/cc) & (g/cc) &  (nm) &  (dyn/cm)  & (erg) & (bar) \\ \hline 
      1:0:0         & 5.66  & 0.95    & 0.66  & 1.30  & 4.9       &  6.6      &  660     \\
      0:1:0         & 3.16  & 1.03    & 0.91  & --    & 4.7       &  1.9      &  870     \\
      0:0:1         & 6.33  & 0.92    & 0.51  & --    & 1.5       &  2.6      &  580     \\ \hline
      7:1:0         & 5.56  & 0.95    & 0.68  & 0.93  & 6.3       &  8.1      &  580     \\
      3:1:0         & 5.29  & 0.95    & 0.68  & 1.16  & 4.4       &  5.2      &  460     \\
      2:1:0         & 5.16  & 0.95    & 0.67  & 1.42  & 5.5       & 6.1       &  500     \\
      1:1:0         & 4.75  & 0.96    & 0.73  & 1.87  & 4.9       & 4.6       &  400     \\ \hline
      7:0:1         & 5.73  & 0.95    & 0.63  & 0.90  & 4.4       & 6.0       &  640     \\
      3:0:1         & 5.81  & 0.95    & 0.64  & 1.10  & 3.7       & 5.2       &  620    \\
      2:0:1         & 5.88  & 0.95    & 0.61  & 0.90  & 4.2       & 6.0       &  570     \\
      1:0:1         & 5.91  & 0.94    & 0.60  & 1.42  & 2.4       & 3.5       &  550     \\ \hline
      1:1:1         & 5.17  & 0.94    & 0.72  & 2.20  & 4.9       & 5.5       &  350     \\
      1:2:1         & 4.99  & 0.96    & 0.80  & 2.52  & 4.5       & 4.7       &  390     \\
      2:1:1         & 5.82  & 0.93    & 0.77  & 3.33  & 2.3       & 3.3       &  350     \\
      5:5:1         & 4.94  & 0.95    & 0.74  & 1.48  & 5.2       & 5.3       &  480     \\
      2:2:1         & 4.89  & 0.94    & 0.69  & 2.50  & 3.9       & 3.9       &  360     \\ \hline
    \end{tabular}
    \caption{Structural properties of the SC lipid bilayers for different
      ratios of CER:CHOL:FFA at 340K.}
    \label{tab.struct}
  \end{center}
\end{table*}

\begin{figure}[htbp]
\centerline{\includegraphics[width=0.48\linewidth,width=0.5\linewidth, clip=]{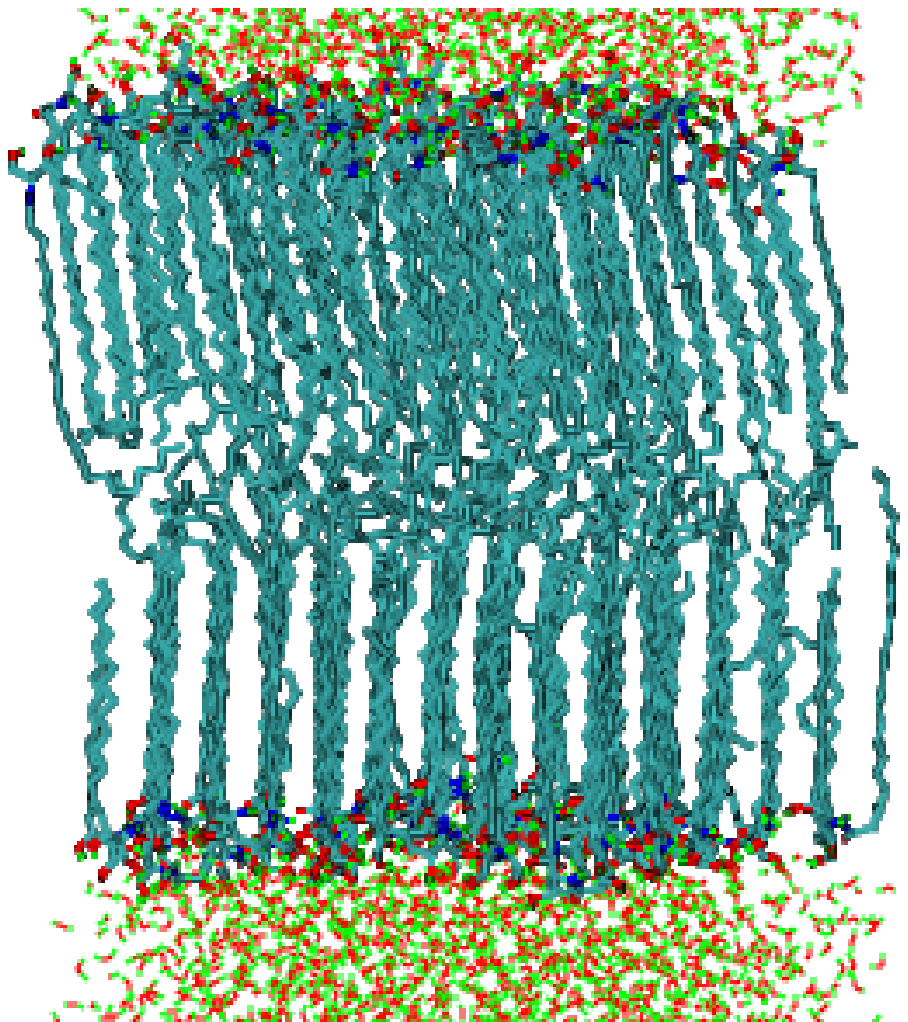} \hfil %
\includegraphics[width=0.48\linewidth,width=0.5\linewidth, clip=]{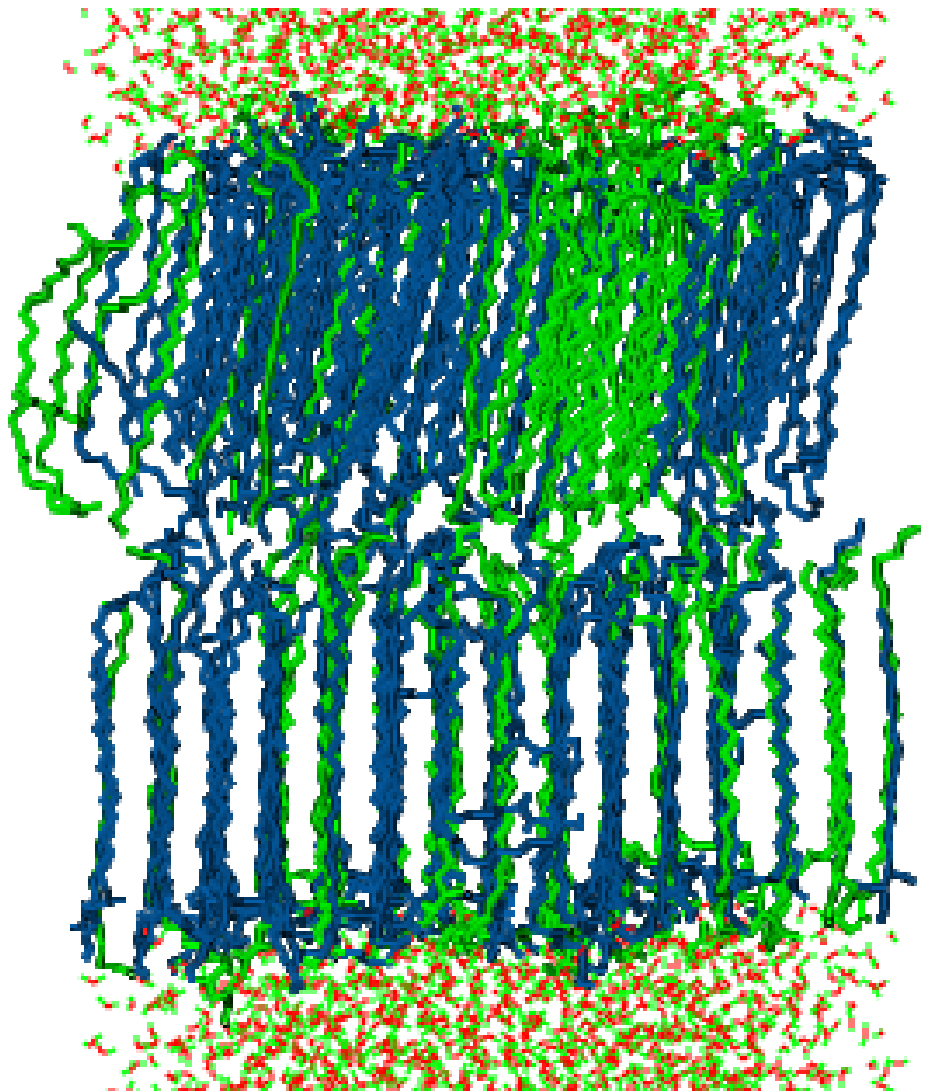} }
\vskip0.2cm
\centerline{(a) \hskip0.6\linewidth (b)}
\vskip0.2cm
\centerline{\includegraphics[width=0.48\linewidth,height=0.5\linewidth, viewport=50 0 450 300,clip=]{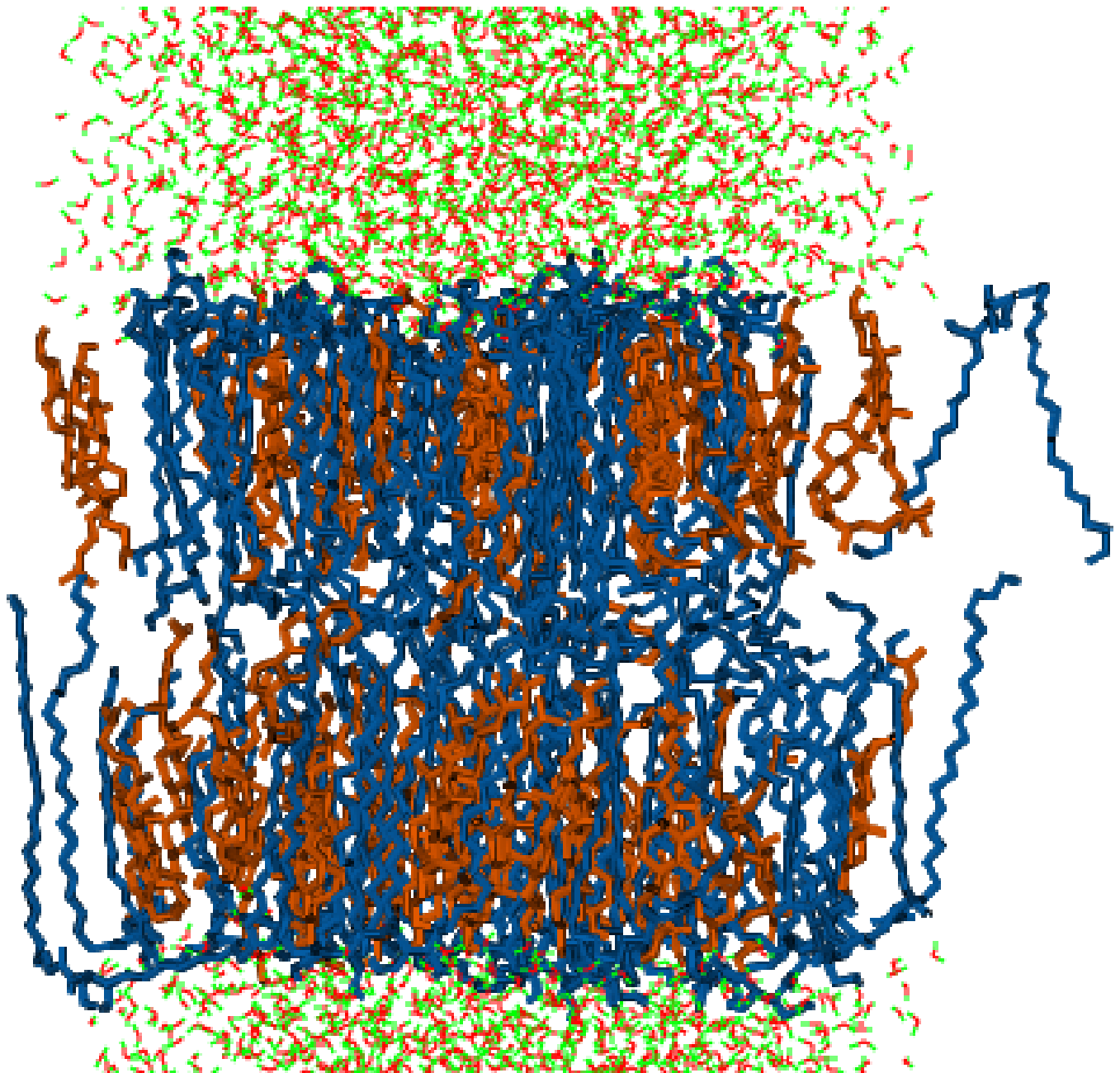} \hfil %
\includegraphics[width=0.48\linewidth,height=0.5\linewidth,viewport=0 0 340 300, clip=]{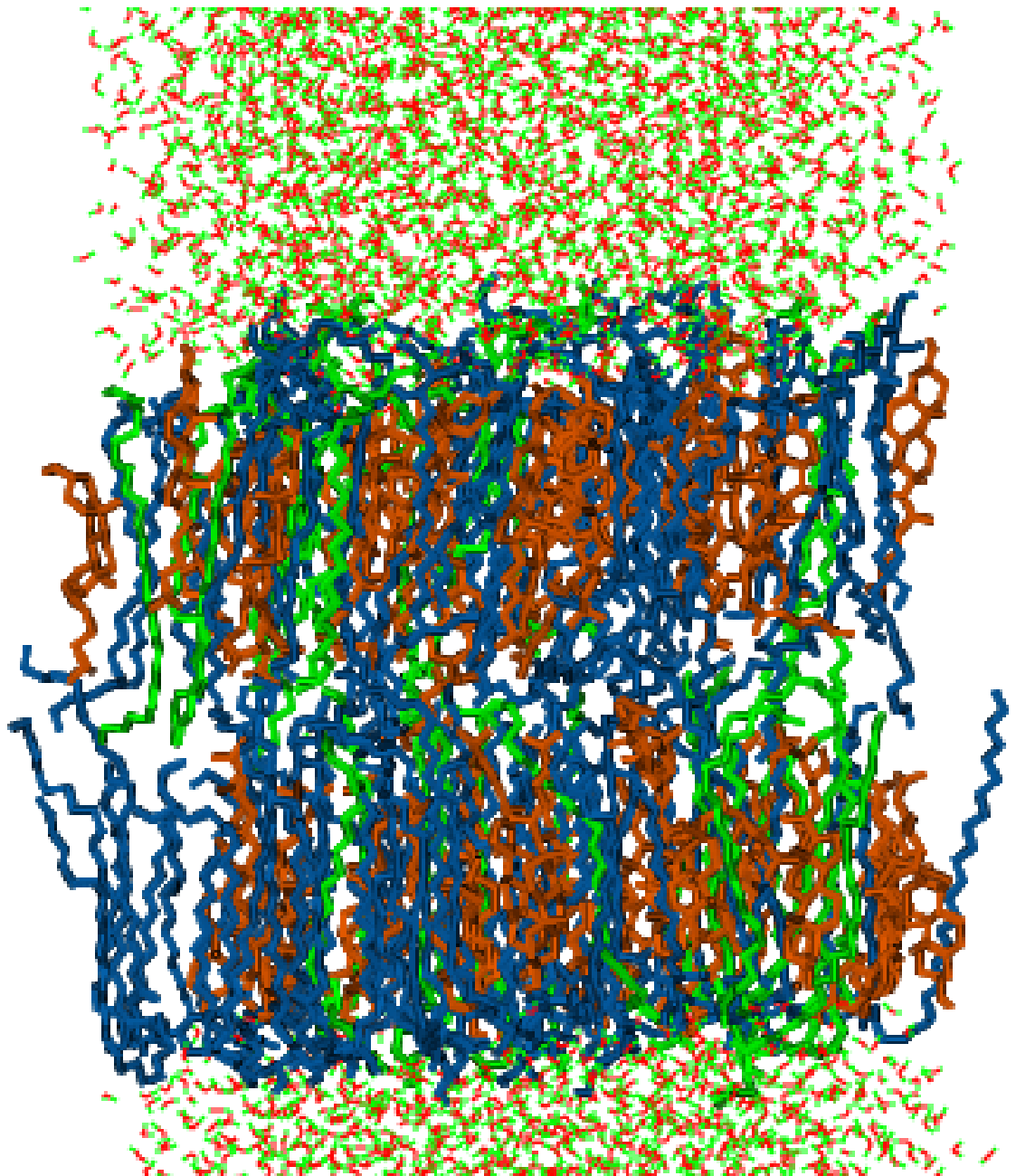} }
\vskip0.2cm
\centerline{(c) \hskip0.6\linewidth (d)}
\caption{Snapshots of (a) pure CER, (b) equimolar CER-FFA mixture
  (1:0:1), (c) equimolar CER-CHOL mixture (1:1:0) and (d) 2:2:1
  mixture of CER, CHOL and FFA at 340K. Only part of the water box is
  shown. CER, CHOL, and, FFA are colored as blue, orange, and green
  respectively. Water molecules are drawn with thin lines.  }
\label{fig.conf.340K}
\end{figure}

Fig.~\ref{fig.conf.340K} shows snapshots (a) for pure CER, (b)
equimolar CER-FFA, (c) equimolar CER-CHOL and (d) 2:2:1 mixture of
CER, CHOL and FFA.  The tails of CER molecules
(Fig.~\ref{fig.conf.340K}a) retain substantial ordering.  Long chain
FFA molecules (Fig.~\ref{fig.conf.340K}b) fall in registry with the
CER molecules. The slightly longer length of the FFA molecules is
accommodated by partially increasing the tail order in the CER
molecules.  The tails in the leaflets arrange in a slightly tilted
orientation with respect to the layer normal. Thus, even though the
amount of ordering is the same on both leaflets, the particular
orientation of Fig.~\ref{fig.conf.340K} highlights the ordering in the
lower leaflets at the cost of obscuring the order in the upper
leaflets.

The head groups of the CHOL (Fig.~\ref{fig.conf.340K}c) mostly stay at
the water-lipid boundary.  The shorter length of the CHOL molecules
squash the bilayer, with the central region comprising primarily CER
tails. The ordering in the CER tails is decreased compared to pure CER
bilayers and the tails from the two leaflets overlap strongly.

The molecular arrangements in the ternary systems
(Fig.~\ref{fig.conf.340K}d) are somewhat intermediate to the binary
CR2-FA and CR2-CHOL mixtures. Single chain FFA molecules are more
flexible than CER molecules, which  induces more FFA atoms in the
midplane disordered phase and  thus increases the tail order of CER molecules
to some degree compared to CER-CHOL mixtures.

\begin{figure}[htbp]
\centerline{\includegraphics[width=0.45\linewidth,clip=]{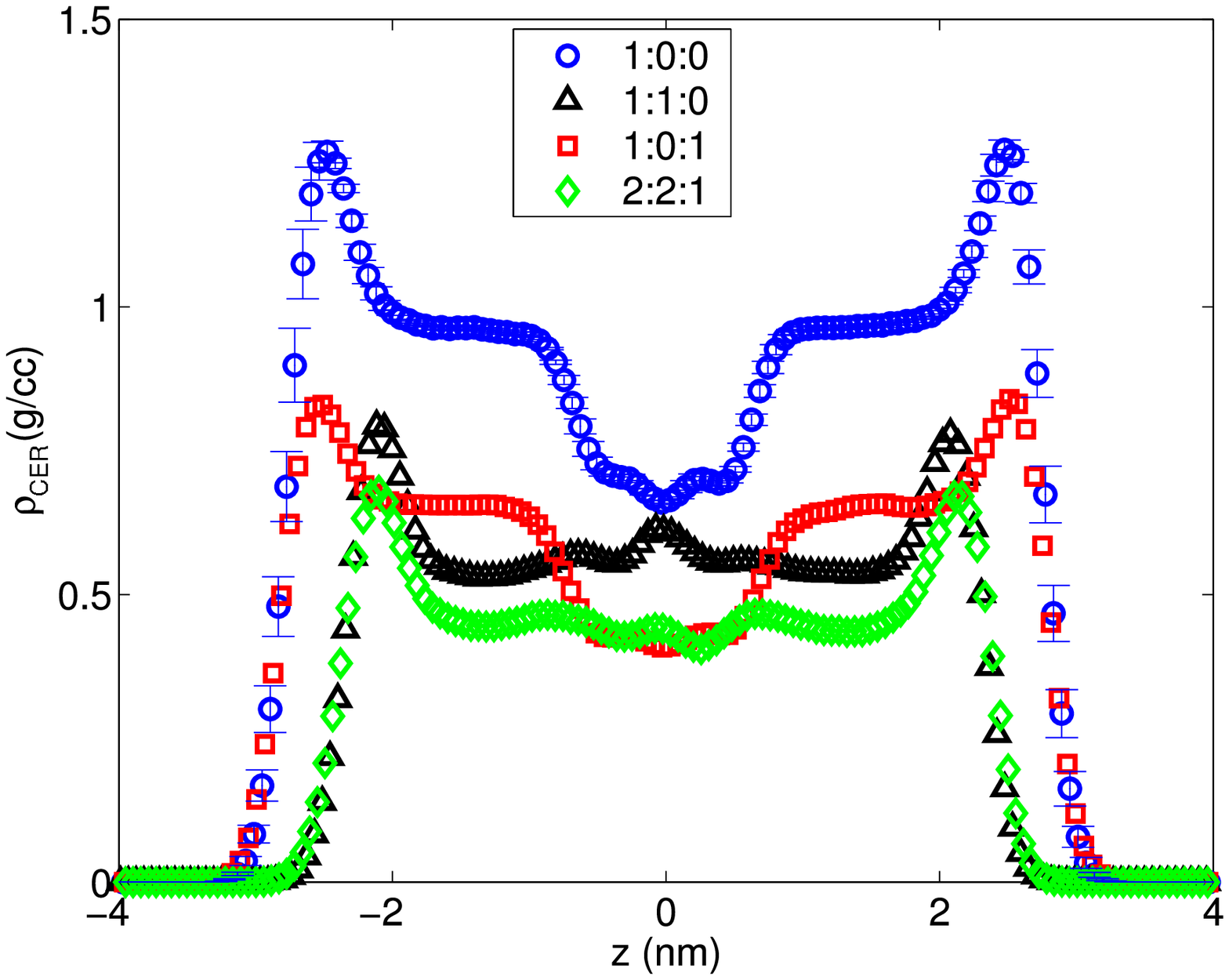} \hfil %
\includegraphics[width=0.45\linewidth,clip=]{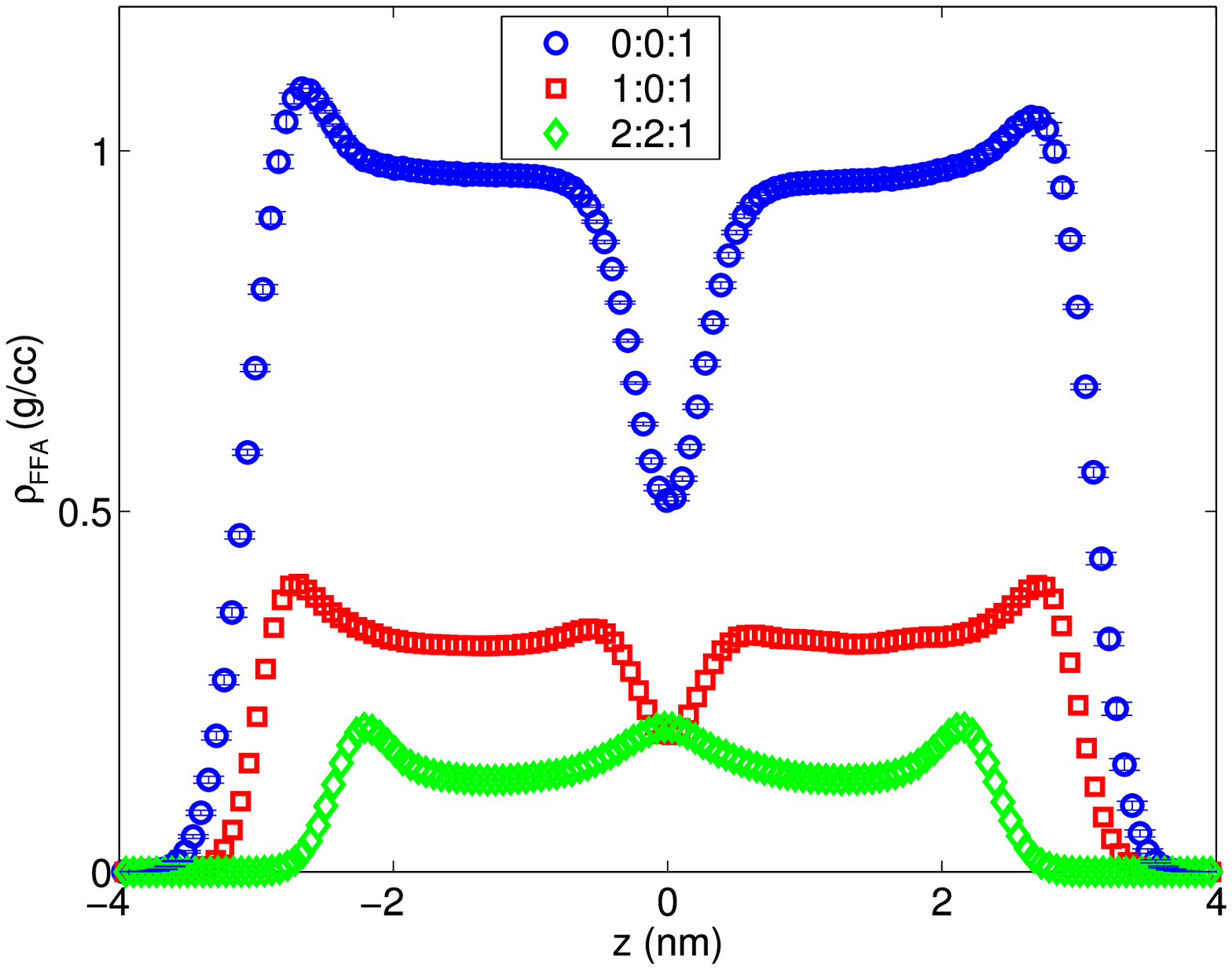} }
\vskip0.2cm
\centerline{(a) \hskip0.6\linewidth (b)}
\vskip0.2cm
\centerline{\includegraphics[width=0.45\linewidth,clip=]{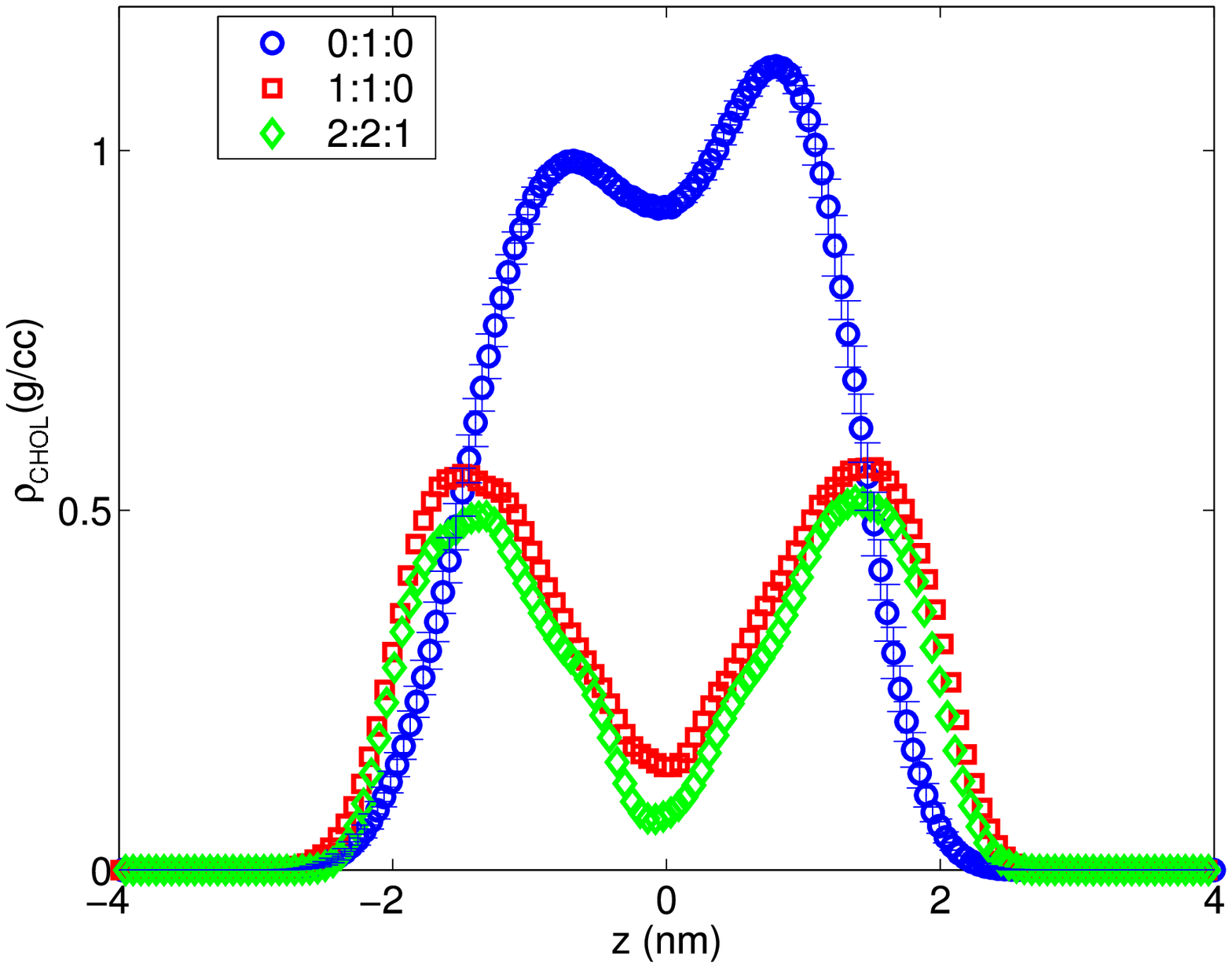} \hfil %
\includegraphics[width=0.45\linewidth,clip=]{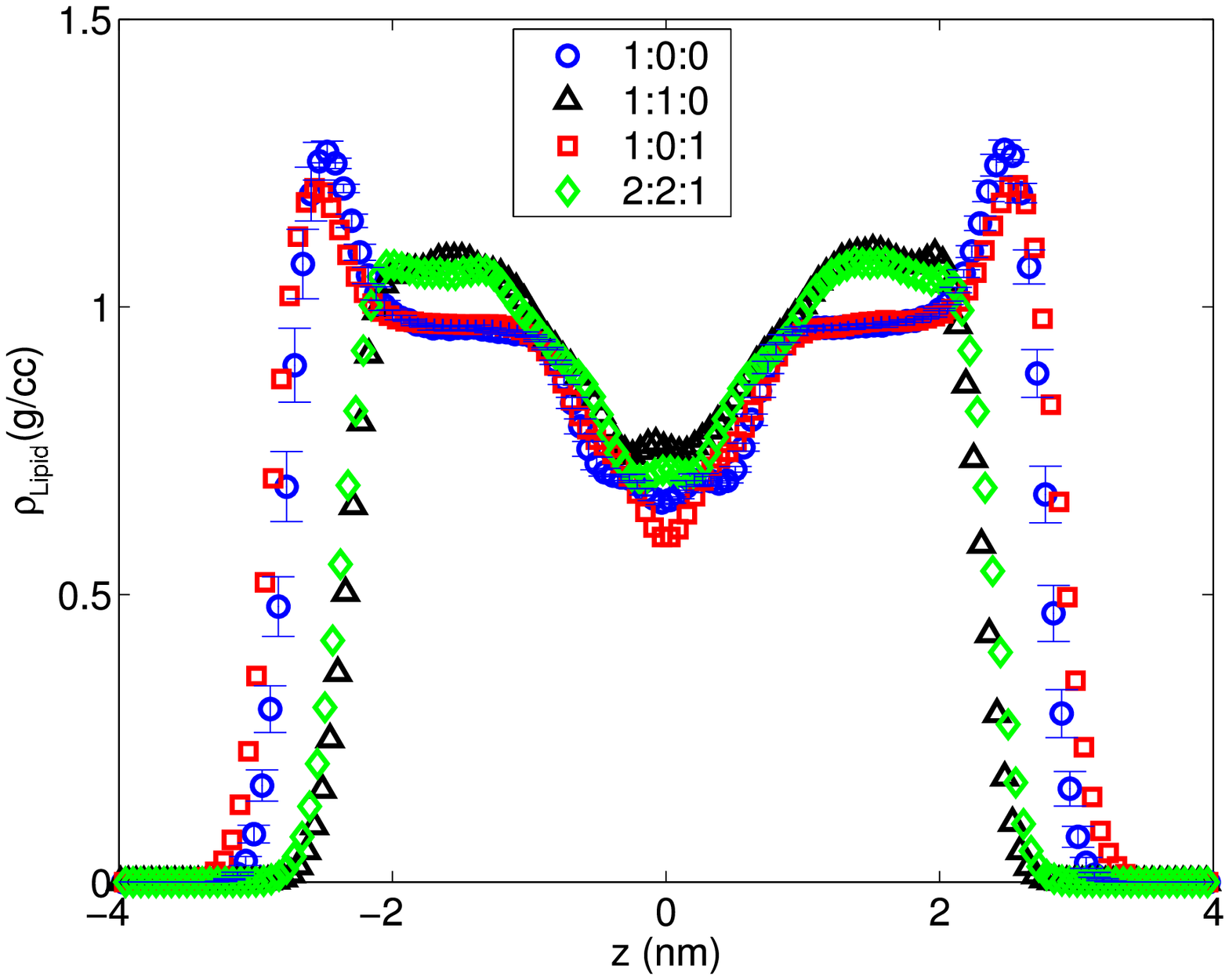} }
\centerline{(c) \hskip0.6\linewidth (d)}
\caption{Lipid densities along the bilayer normal directions at 340K
  for (a) CER (b) FFA (c) CHOL and (d) all lipid molecules considered
  together. Legends show molar ratio of CER:CHOL:FFA.}
\label{fig.lden.340K}
\end{figure}

\begin{figure}[htbp]
\centerline{ \includegraphics[width=3.5in, viewport= 0 0 400
  280,clip=true]{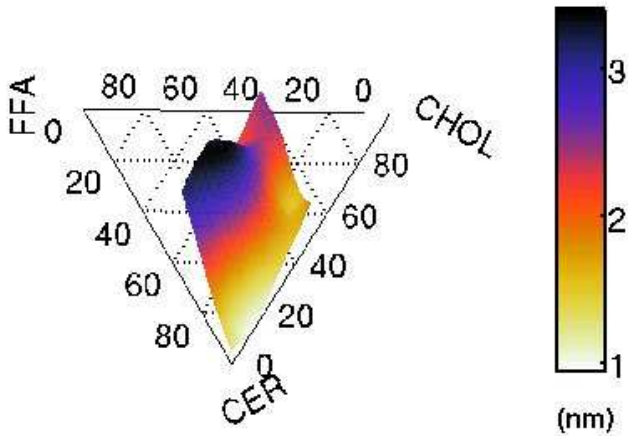} } 
\caption{Estimate of overlap $\lambda_{ov}$ of the CER tails from the
  two leaflets (Eq.~\ref{eq.overlap}) at 340K.}
\label{fig.lambdaoverlap}
\end{figure}

To investigate the arrangement of the different components further, in
Fig.~\ref{fig.lden.340K} we plot the density of the lipid components
along the $z$ direction for different compositions.
Fig.~\ref{fig.lden.340K}a shows the density profile of the CER atoms.
Both pure CER and CER-FFA systems have a high density near the head
groups, followed by a constant density region due to the ordered tail
atoms and then a region of lower density at the midplane, which covers
almost 2\,nm. For CER-CHOL and three component mixtures, the density from
the tail atoms of the CER molecules remains almost constant throughout
the bilayer. In fact, for CER-CHOL mixtures, there is an increase in
the local CER density at the bilayer midplane due to increased interdigitation
of the molecules from the two leaflets. The plateau densities in
Fig.~\ref{fig.lden.340K} are fixed by the relative abundance of the
different molecules in the mixtures.  Fig.~\ref{fig.lden.340K}b shows
the density profile of the FFA atoms. For pure FFA and CER-FFA
mixtures, the density profiles are qualitatively similar to that of
pure CER bilayers, except the region of low density inter-leaflet
space is narrower. In the 3-component mixtures, the density minimum is
replaced by a local density maximum because of inter-leaflet overlaps.
The density profile of the CHOL molecules (Fig.~\ref{fig.lden.340K}c)
does not have the constant density hydrocarbon regions present for the
FFA and CER molecules. The profile for the pure CHOL bilayer is not
symmetric and the interleaflet density quite high, signifying
significant transfer between the two leaflets during the simulation
time scale.
Fig.~\ref{fig.lden.340K}d shows the total lipid densities. The effect
of fatty acid is to increase the bilayer thickness and reduce the
density by a small amount. CHOL, when present, reduces the bilayer
thickness and increases the density at the tail region.

The overlap of the lipids from the two leaflets (partial
interdigitation) is expected to increase the inter-leaflet friction
and couple the dynamics of the two leaflets.
Fig.~\ref{fig.lambdaoverlap} shows a surface plot of $\lambda_{ov}$,
a measure of interdigitation, as a function of composition. The three
component mixture is characterized by large $\lambda_{ov}$. One of the
SC lipids, not considered in this study, ceramide~1 (ceramide EOS)
contains an additional long-chain $\omega$-hydroxy acid (with number
of carbon atoms $>$ 30) linked to the fatty acid tail. This
effectively makes the length of the long tail of ceramide~1 nearly
double to that of the other members of the ceramide family.  The
presence of ceramide~1 will probably introduce significantly more
interdigitation and inter-leaflet coupling.

\begin{figure}[htbp]
\centerline{ \includegraphics[width=3.5in, clip=true]{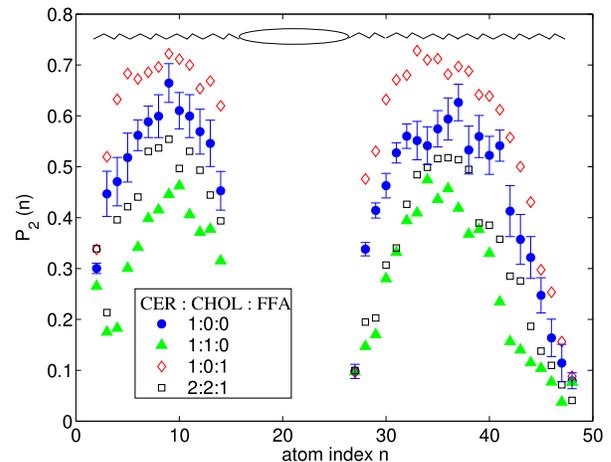} }
\caption{Local tail order parameter $P_2(n)$ as a function of atom
  index $n$ on a CER molecule (Fig.\ref{fig.scskeletal}) for various
  compositions at 340K.   Errorbars are shown for the pure CER
  data (circles).}
\label{fig.tailop}
\end{figure}

\begin{figure}[htbp]
\centerline{ \includegraphics[width=3.5in, clip=true]{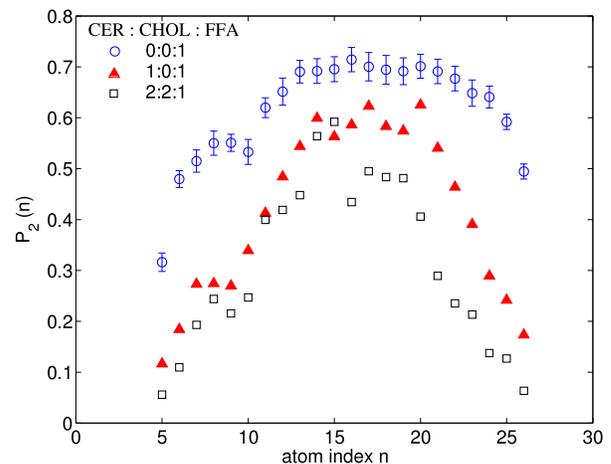} }
\caption{Local tail order parameter $P_2(n)$ as a function of atom
  index $n$ on a FFA molecule (Fig.\ref{fig.scskeletal}) for various
  compositions at 340K.}
\label{fig.tailopfa}
\end{figure}
Fig.~\ref{fig.tailop} shows the tail order parameter of CER atoms at
340K for different composition ratios.  The presence of CHOL reduces
the nematic order, while FFA increases the order. 
The presence of
either CER or CHOL reduces the order of the FFA atoms compared to a
pure FFA bilayer (Fig.~\ref{fig.tailopfa}). 
This differs from
phospholipid membranes, where the planar shape of cholesterol
typically increases the nematic order of the phospholipid tails. When
mixed with the long ceramides, the shorter cholesterol molecules tend
to encourage a thinner membrane by disordering the longer ceramide
tails so that they can fill the space around the cholesterol. This
also accounts for the increased overlap of the CER tails upon adding
cholesterol, as depicted in Fig.~\ref{fig.lambdaoverlap}.

\begin{figure}[htbp]
\centerline{ \includegraphics[width=3.2in, clip=true]{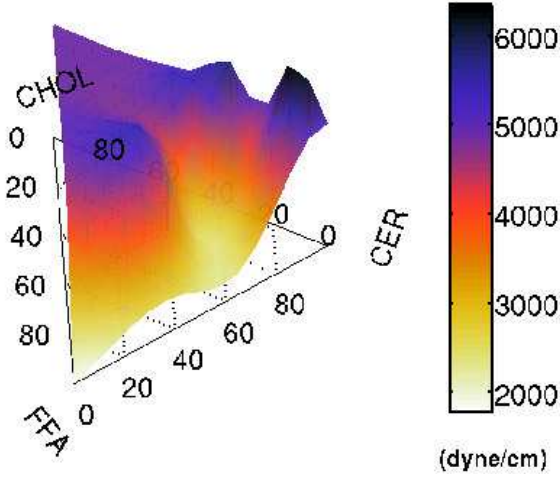} }
\caption{Area compressibility modulus $\kappa_A$ of the bilayers at 340K.}
\label{fig.kappaa}
\end{figure}

\begin{figure}[htbp]
\centerline{
\includegraphics[width=0.48\linewidth, clip=true]{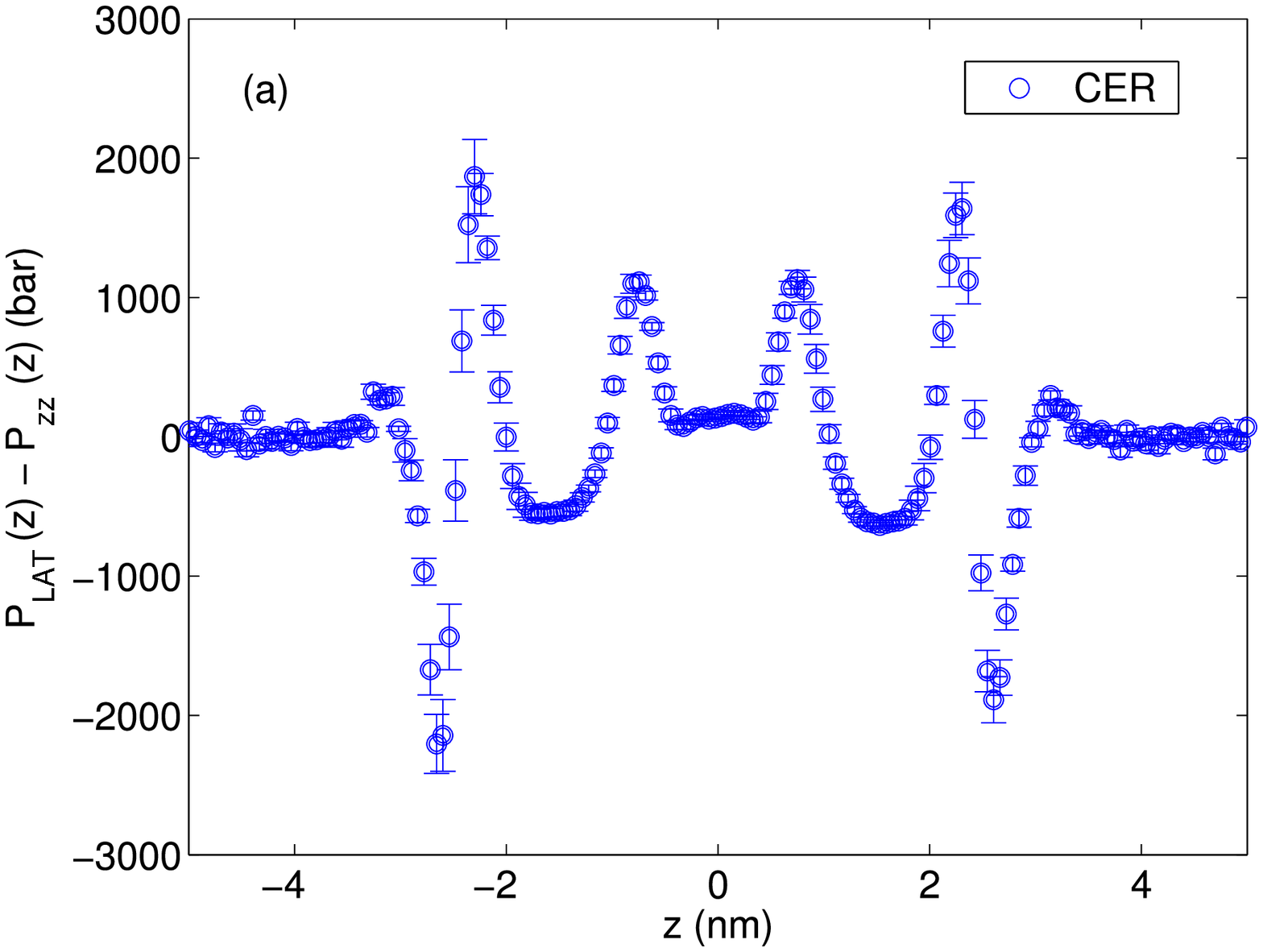} \hfill
\includegraphics[width=0.48\linewidth, clip=true]{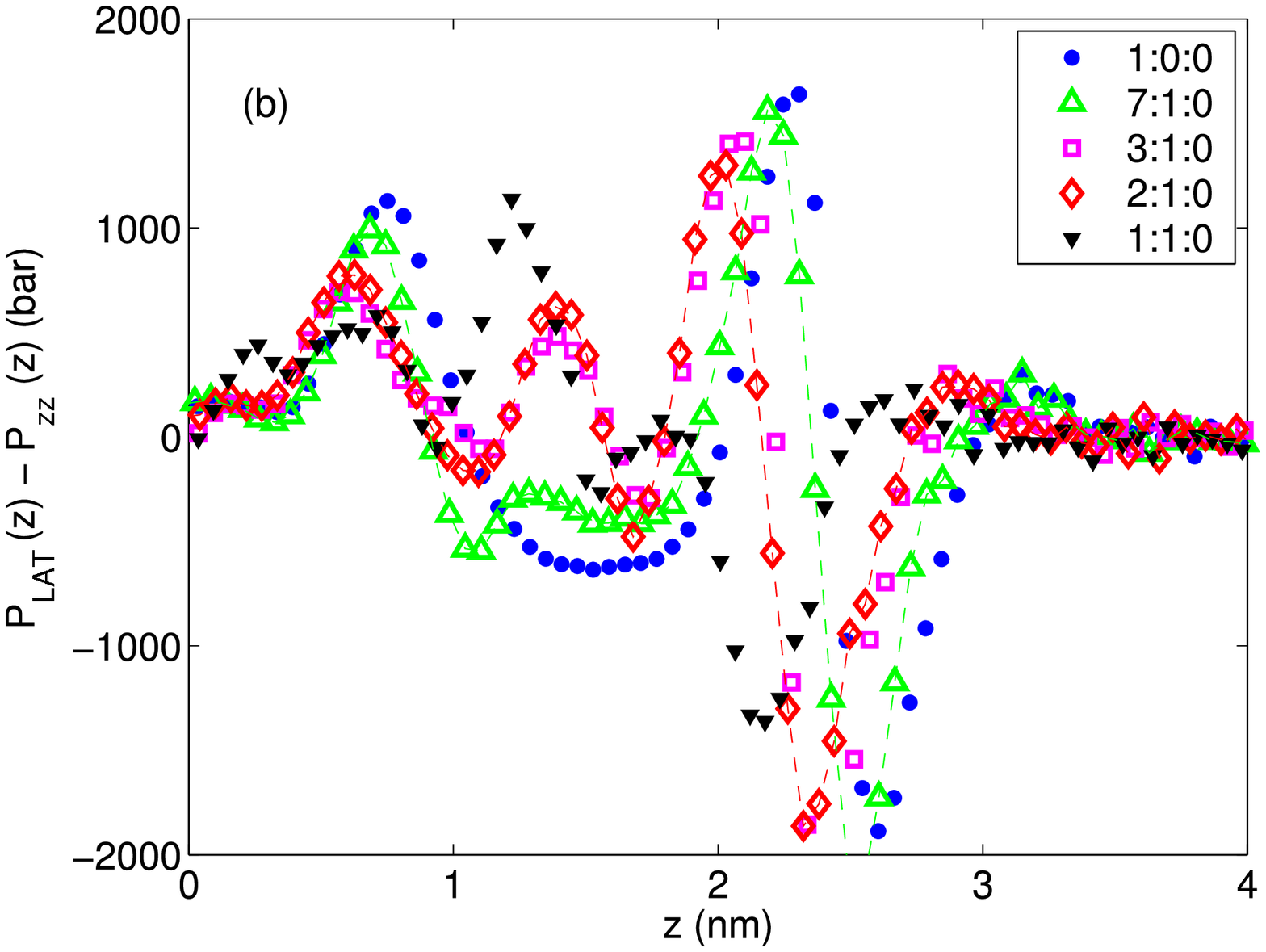} }
\vskip0.2cm
\centerline{
\includegraphics[width=0.48\linewidth, clip=true]{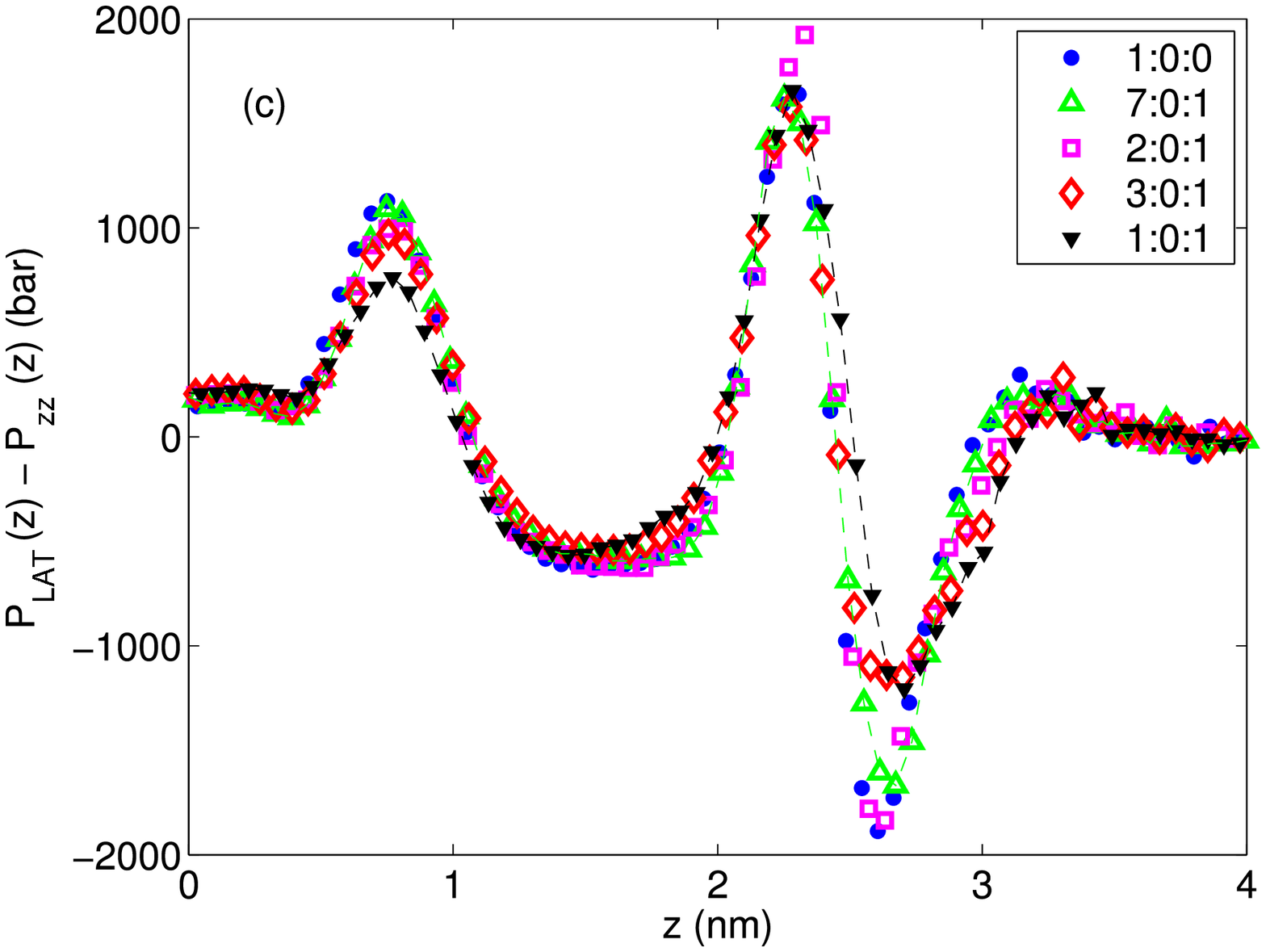} \hfill
\includegraphics[width=0.48\linewidth, clip=true]{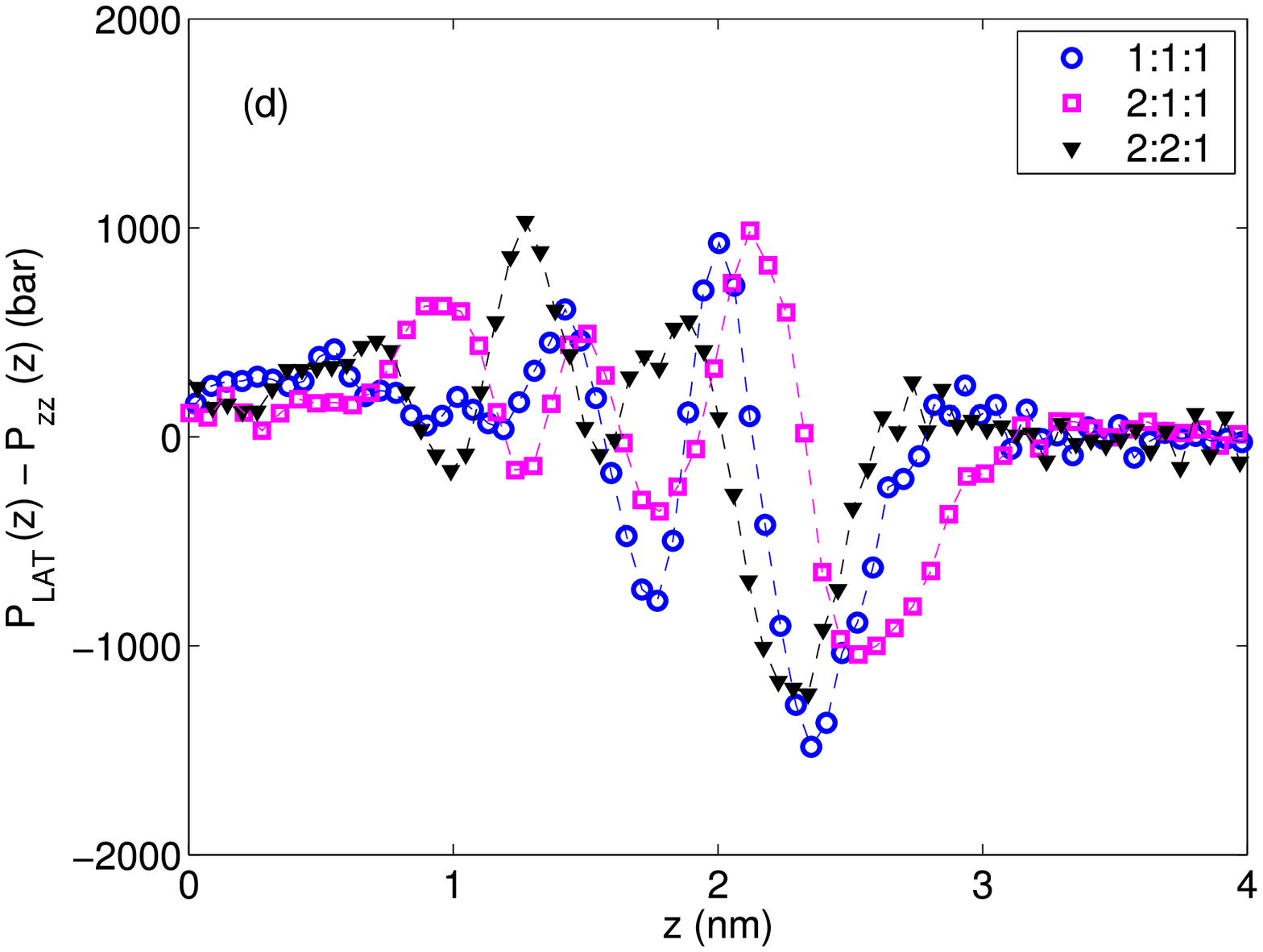} }
\caption{Difference between the lateral and perpendicular components
  of the local pressure ($\delta P = P_{LAT} - P_{zz}$) as a function
  of distance from the bilayer mid-plane at 340K. (a) $\delta P (z)$
  for pure CER.  (b) and (c) respectively show the effect of adding
  CHOL and FFA to CER bilayers. Because the local pressure profile is
  symmetric about the bilayer midplane, only one side of the data is
  reproduced for clarity. (d) $\delta P$ for selected three component
  mixtures.  For some data sets, smooth lines joining the points are
  drawn as a guide to eye.  Legends show the ratio of CER:CHOL:FFA.  }
\label{fig.localp} 
\end{figure}

Fig.~\ref{fig.kappaa} shows a surface plot of the area compressibility
$\kappa_A$ on a ternary diagram of the three components. The bending
modulus $\kappa$, calculated from the polymer brush theory 
\cite{rawicz.polybrush.00} using
$\kappa_A$ and the bilayer thickness, behaves in a similar fashion.
Close to the skin composition CER:FFA:CHOL=2:2:1 the bilayer becomes
softer, with comparatively smaller $\kappa_A$ and $\kappa$. The
absolute magnitude of the elastic constants remain much higher than in
the fluid phase of phospholipid bilayers.

\begin{figure}[htpb]
\centerline{
\includegraphics[width=3.5in, clip=true]{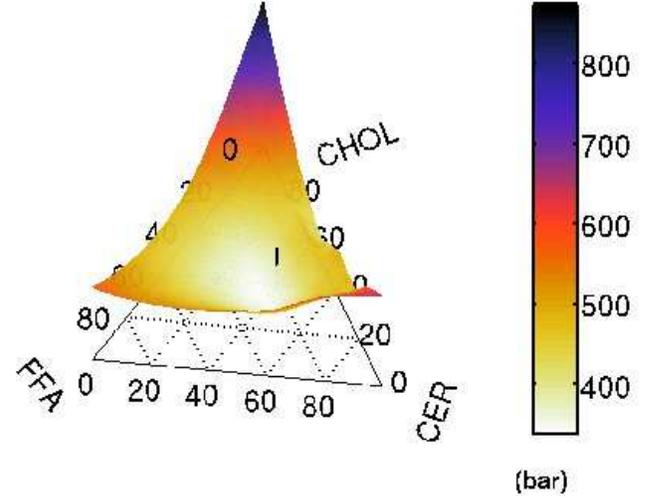} }
\caption{Local stress $\bar{\epsilon}_P$ (Eq.~\ref{eq.localp.en}) in
  the bilayers associated with the difference in the lateral and the
  perpendicular components of the pressure tensor. 2:2:1 composition
  is indicated with a vertical line.}
\label{fig.localp.en}
\end{figure}

Fig.~\ref{fig.localp} shows the difference between the lateral and the
normal pressure $\delta P(z)$ along the bilayer normal direction.
Inside the bilayers, pure CER (Fig.~\ref{fig.localp}a) shows large
variations in $\delta P(z)$ as a function of $z$. Similar but less
pronounced variations in $\delta P(z)$ are observed in phospholipid
membranes.  For example, \citet{lindahl.00} found a maximum variation
of $|\delta P(z)| < 500$~bar for a DPPC bilayer at 323K, which is 5
times smaller than the present case.  The addition of CHOL
(Fig.~\ref{fig.localp}b) introduces additional peaks in $\delta P(z)$.
FFA reduces the peak height marginally (Fig.~\ref{fig.localp}c).  The
three component mixtures (\ref{fig.localp}d) show much less pronounced
variations in $\delta P(z)$ than the pure ceramide or any of the 2
component mixtures. A large $\delta P$ can be interpreted as a local
moment acting on the molecules, making a deviation from the flat
interface of bilayer more likely.  The magnitude of $\delta P$ also
can be viewed as a local stress $\bar{\epsilon}_P$.
Fig.~\ref{fig.localp.en} shows a ternary plot of $\bar{\epsilon}_P$
defined through Eq.~\ref{eq.localp.en}.  The vertical line shows the
position of composition 2:2:1, where $\bar{\epsilon}_P$ shows a
minimum. To gain some insight into the energy scale, we note that
$10^3\,\textrm{bar}\, \sim 0.7 k_B T$ per methyl group, while the
trans-gauche energy difference is $\sim 2 k_B T$ per dihedral bond.
Thus, in the absence of prohibitive energy barriers, the system can
introduce gauche defects to reduce the local variation of $\delta P$.

\section{Conclusions}

We have presented molecular dynamics simulation results for various
composition ratios of the three main constituent components of stratum
corneum lipid layers, namely, CER~NS~24:0, FFA~24:0 and CHOL.  The
long asymmetric tails of the CER molecules form a dense bilayer phase
in water, with considerable interdigitation of the tails from the two
leaflets and strong nematic order. CHOL, being rigid and smaller in
length compared to CER, acts as a molecular clamp, squashing the
bilayer and increasing the in-plane density. The resulting structure
shows large lateral stress variations, which is relieved by the
presence of FFA. The three component mixtures are characterized by a
higher
density and comparatively smaller area compressibilities and bending
moduli.

The lipids in real skin stratum corneum have large polydispersity in
tail length and have different head groups.  This work does not
address the effect of polydispersity. We also study bilayers in excess
water, which is quite different from limited water environment in the
skin. The strong local pressure fluctuations in the bilayer suggests
that the bilayer may not be the most stable structure and it will be
interesting to simulate the lipids in limited water conditions. The
hydrocarbon densities in the bilayers were found to be quite large
compared to phospholipids. This is probably responsible for three
orders of magnitude smaller permeability of water through stratum
corneum as compared to plasma membranes.
\appendix
\section{Equilibration time}
\begin{figure}[htbp]
\centerline{\includegraphics[width=3.25in,clip=true]{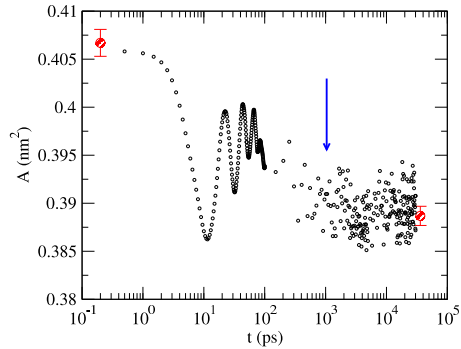}}
\caption{Equilibration of area/lipid for a CER bilayer after a 
sudden change in thermostat temperature from 360 K to 300 K. 
The big circles with error-bars are the values obtained from different 
runs at 360K and 300K.}
\label{fig.app.jump} 
\end{figure}

To probe the typical time it takes for the system to adapt the 
equilibrium conformation for a given temperature, we used an equilibrated 
configuration of 128 CER lipids (1:0:0 composition) at 360 K and monitored 
the instantaneous area/lipid as a function of time 
(small black circles in Fig~\ref{fig.app.jump}) after abruptly changing the 
desired thermostat temperature to 300 K. We also show the 
area/lipid calculated from separate runs at 360 K and at 300 K as 
shaded big circles with error bars at the two extreme ends of the data. 
As can be seen from the plot, the area/lipid adapts to the new temperature 
with a time scale of around 1 ns (position of the arrow). 
In the paper we use equilibration times of 10 ns for each 
10K temperature shift - which gives ample time for the molecules 
to reorganize themselves into the equilibrium conformation corresponding 
to the set temperature.

\section{Effect of long range electrostatics and system size}
To explore the dependence of measured quantities on system size and run times, 
we have carried out some additional simulations on pure CER bilayers 
(at $300\,\textrm{K}$) and 
bilayers with 2:2:1 composition ratio of CER, CHOL, FFA 
(at $340\,\textrm{K}$).
These simulations 
use much larger number of lipid molecules and much longer run times 
than the simulations reported in the paper. For these simulations,
we use both the group-based cut-off and PME methods to probe the
effect of long-range electrostatics. In simulation with PME, the grid spacing 
was chosen to be 0.1 nm and fourth order polynomial interpolation was used. 
A summary of the number of
molecules used and timescales probed is given 
in Table.~\ref{tab.app.syssize}. To get the initial configuration for a larger 
system, we use the already equilibrated configuration at the required
temperature from the simulations presented in the main paper and replicate 
it in the x and y directions. To reduce disk space requirements, in most 
cases of these set of simulations, we save configurations only at intervals 
of 0.2 ns (as opposed to 0.5 ps in the main paper). Thus we expect larger 
statistical errors in the data reported in this appendix as compared to the 
main paper.

\begin{table}[htbp]
\begin{tabular}{|c|l|l|l|l|l|l|}\hline
Composition & \multicolumn{4}{|c|}{Number of molecules}
& \multicolumn{2}{|c|}{Run time (ns)} \\ \hline
CER:CHOL:FFA & &&& && \\
(molar ratio) &CER &CHOL & FFA & SOL & Cutoff & PME \\ \hline
2:2:1  &56 & 56 &32 & 5250 & 100 & 100 \\
(340 K) & 224 & 224 & 128 & 21000 & 70 & 40 \\
 & 504 & 504 & 288 & 47250 & 50 & 16 \\ \hline
1:0:0 &	128 & -- & -- & 5250 & 30 & 30 \\ 
(300 K) & 512 & -- & -- & 21000 & 30 & 16 \\
 & 1152	& -- & -- & 47250 & 20 & 10 \\ \hline
\end{tabular}
\caption{Number of molecules and simulation times used to probe effect of
long range electrostatics and system size.}
\label{tab.app.syssize}
\end{table}

In Table~\ref{tab.app.pme}, we report the energies (normalized by the number 
of lipid molecules in the system) for 2:2:1 system calculated separately 
with group based cut-off and particle mesh Ewald summation (PME).
The two important observations from the energy values are that 
(i) the difference between the group based cut-off and PME schemes in the 
total energy is about 2\% 
and (ii) the normalized energies are independent of the system sizes. 
Both of these two observations show that the interactions with the 
periodic images do not contribute significantly in the electrostatic energy.  
Table~\ref{tab.app.pme} also shows the area compressibility and the 
average bilayer thickness calculated from these simulations. The effect of 
the system size and handling of electrostatics do not affect the values 
significantly.

\begin{table}[htbp]
\begin{tabular}{|c|l|l|l|l|l|l|}\hline
\multicolumn{2}{|c|}{Number of} & Coulomb & Coulomb &
Total & $\kappa_{A}\,^{\dagger}$ & 2 d  \\
\multicolumn{2}{|c|}{lipids} & (SR)$^{\ast}$ & (LR)$^{\ast}$ & energy$^{\ast}$ &
 & (nm) \\ \hline 

144 & cut-off & -1.7675 & --- & -1.1310 & 2.6 (2) & 4.90 (2) \\
    &PME & -1.6363 & -0.1101 & -1.1114 &2.0 (3) & 4.95 (2) \\ \hline
576 & cut-off & -1.7676 &--- &	-1.1308 & 2.5 (5) & 4.97 (2) \\
 & PME & -1.6364 & -0.1101 & -1.1105 & 3.6 (3) & 4.91 (2) \\ \hline
1296 &cut-off & -1.7676 & ---& -1.1302	& 2.3 (5) & 4.93 (2) \\
 & PME& -1.6364	& -0.1102 & -1.1104 & 2.9 (4) & 4.90 (1) \\ \hline
\end{tabular}
\caption{Short range (SR) and long range (LR) electrostatic energies, 
total energy, area compressibility ($\kappa_A$) and 
bilayer thickness ($2d$) for bilayers containing 2:2:1 molar ratio of 
CER2, CHOL and FFA at 340K. Statistical errors of the mean values are 
indicated in the brackets as the uncertainty on the last digit. 
($^{\ast}$ Energies are in MJ/mol and are normalized by the number of 
lipid molecules in the system. 
$^{\dagger}$ $\kappa_A$ is in $10^3\,\textrm{dyn/cm}$.)} 
\label{tab.app.pme}
\end{table}
Table~\ref{tab.app.pme.cr2} shows the effect of system size and method of 
calculating electrostatic interaction for pure CER bilayers at 300K. The 
conclusions about insensitivity of the results on both the system size and 
the long range electrostatic interaction in the case of 2:2:1 lipid mixture 
remains equally valid for the case of CER bilayers.

\begin{table}[htbp]
\begin{tabular}{|c|l|l|l|l|l|l|}\hline
\multicolumn{2}{|c|}{Number of} & Coulomb & Coulomb &
Total & $\kappa_{A}\,^{\dagger}$ & 2 d  \\
\multicolumn{2}{|c|}{lipids} & (SR)$^{\ast}$ & (LR)$^{\ast}$ & energy$^{\ast}$ &
 & (nm) \\ \hline 

128 & cut-off & -2.2045 & ---     & -1.5516 & 7.0 (6) & 5.67 (1) \\
         &PME & -1.9974 & -0.1755 & -1.5272 & 6.1 (6) & 5.68 (1) \\ \hline
512 & cut-off & -2.1996 &---      & -1.5520 & 6.2 (9) & 5.70 (1) \\
        & PME & -1.9973 & -0.1754 & -1.5269 & 7.4 (9) & 5.70 (1) \\ \hline
1152 &cut-off & -2.1994 & ---     & -1.5516 & 5.8 (7) & 5.72 (1) \\
         & PME& -1.9946	& -0.1753 & -1.5245 & 7.4 (9) & 5.72 (1) \\ \hline
\end{tabular}
\caption{Short range (SR) and long range (LR) electrostatic energies, 
total energy, area compressibility ($\kappa_A$) and 
bilayer thickness ($2d$) for pure CER bilayers 
at 300K. Statistical errors of the mean values are 
indicated in the brackets as the uncertainty on the last digit. 
($^{\ast}$ Energies are in MJ/mol and are normalized by the number of 
lipid molecules in the system. 
$^{\dagger}$ $\kappa_A$ is in $10^3\,\textrm{dyn/cm}$.)} 
\label{tab.app.pme.cr2}
\end{table}

For the smallest system sizes used in this study, typical CPU requirements 
for PME based calculations are more than twice as compared to the CPU 
requirement for group-based cut-off. Hence, for the simulation results 
presented in the main paper, we confine ourselves to group-based cut-off 
scheme only.

\begin{acknowledgments}
This work was supported by Yorkshire Forward through the grant YFRID
Award B/302. CD acknowledges SoftComp EU Network of Excellence for
financial support and computational resources. The authors thank
Jamshed Anwar, Simon Connell, Brett Donovan, Andrea Ferrante, Robert
Marriott, Rebecca Notman, Khizar Sheikh, Sathish Sukumaran,
and Barry Stidder for useful
discussions.  
\end{acknowledgments}

\end{document}